\documentclass[journal]{IEEEtran}
\usepackage{amssymb}
\usepackage{graphicx}
\usepackage{epstopdf}
\usepackage{gensymb}
\usepackage{color}
\usepackage{amsmath}
\usepackage{bm}
\usepackage{etoolbox}
\usepackage{cite}
\usepackage{array}
\usepackage{booktabs}
\usepackage{multirow}
\usepackage{comment}

\usepackage{algorithmicx}
\usepackage[ruled]{algorithm}
\usepackage{algpseudocode}
\alglanguage{pseudocode}

\DeclareMathOperator*{\argmin}{\arg\!\min}
\usepackage{threeparttable}
\usepackage{mathtools}
\usepackage{hyperref}
\usepackage{pifont}

\DeclareMathOperator*{\argmax}{argmax}
\usepackage{url}
\urldef{\mailsa}\path|{alfred.hofmann, ursula.barth, ingrid.haas, frank.holzwarth,|
	\urldef{\mailsb}\path|anna.kramer, leonie.kunz, christine.reiss, nicole.sator,|
	\urldef{\mailsc}\path|erika.siebert-cole, peter.strasser, lncs}@springer.com|    

\usepackage[utf8]{inputenc}
\usepackage[english]{babel}

\usepackage{amsthm}

\theoremstyle{definition}

\definecolor{R}{RGB}{0,0,150}

\theoremstyle{remark}

\ifCLASSINFOpdf

\begin{document}

\title{STRIP: A Defence Against Trojan Attacks on Deep Neural Networks}
%
%
%

\author{
	Yansong Gao, Chang Xu, Derui Wang, Shiping Chen, Damith C.~Ranasinghe, and Surya Nepal
	\thanks{Y.~Gao is with the School of Computer Science and Engineering, Nanjing University of Science and Technology, Nanjing, China and Data61, CSIRO, Sydney, Australia. e-mail: yansong.gao@njust.edu.cn}
	\thanks{C.~Xu, S.~Chen, S.~Nepal are with Data61, CSIRO, Sydney, Australia. e-mail: \{chang.xu; shiping.chen; surya.nepal\}@data61.csiro.au.}
	\thanks{D.~Wang is with the School of Information Technology, Deakin University, Burwood and Data61, CSIRO, Australia. e-mail: derekw@deakin.edu.au.}
	\thanks{D.~C. Ranasinghe is with School of Information Technology, Deakin University, Australia and Data61, CSIRO, Australia. e-mail: derekw@deakin.edu.au.}
	\thanks{Cite as: Yansong Gao, Change Xu, Derui Wang, Shiping Chen, Damith C. Ranasinghe, and Surya Nepal. 2019. STRIP: A Defence Against Trojan Attacks on Deep Neural Networks. In 2019 Annual Computer Security Applications Conference (ACSAC ’19), December 9–13, 2019, San Juan, PR, USA. ACM, New York, NY, USA, 13 pages. https://doi.org/10.1145/3359789.3359790}
}


\maketitle
\begin{abstract}
A recent trojan attack on deep neural network (DNN) models is one insidious variant of data poisoning attacks. Trojan attacks exploit an effective \textit{backdoor} created in a DNN model by leveraging the difficulty in interpretability of the learned model to misclassify {\it any inputs} signed with the attacker's chosen trojan trigger. Since the trojan trigger is a secret guarded and exploited by the attacker, detecting such {\it trojan inputs} is a challenge, especially at run-time when models are in active operation. This work builds {\underline {STR}}ong {\underline I}ntentional {\underline P}erturbation (STRIP) based \textit{run-time} trojan attack detection system and focuses on vision system. We intentionally perturb the incoming input, for instance by superimposing various image patterns, and observe the randomness of predicted classes for perturbed inputs from a given deployed model---malicious or benign. A low entropy in predicted classes violates the input-dependence property of a benign model and implies the presence of a malicious input---a characteristic of a trojaned input. The high efficacy of our method is validated through case studies on three popular and contrasting datasets: MNIST, CIFAR10 and GTSRB. We achieve an overall false acceptance rate (FAR) of less than 1\%, given a preset false rejection rate (FRR) of 1\%, for different types of triggers. Using CIFAR10 and GTSRB, we have empirically achieved result of 0\% for both FRR and FAR. We have also evaluated STRIP robustness against a number of trojan attack variants and adaptive attacks.
\end{abstract}

\IEEEpeerreviewmaketitle
\section{Introduction}
Machine learning (ML) models are increasingly deployed to make decisions on our behalf on various (mission-critical) tasks such as computer vision, disease diagnosis, financial fraud detection, defending against malware and cyber-attacks, access control, surveillance and so on~\cite{lecun2015deep,wang2017adversary,tang2016deep}. However, the safety of ML system deployments has now been recognized as a realistic security concern~\cite{stoica2017berkeley,guo2018lemna}. In particular, ML models can be trained (e.g., outsourcing) and provided (e.g., pretrained model) by third party. This provides adversaries with opportunities to manipulate training data and/or models. Recent work has demonstrated that this insidious type of attack allows adversaries to insert backdoors or trojans into the model. The resulting trojaned model~\cite{chen2017targeted,ji2018model,gu2017badnets,zou2018potrojan,bagdasaryan2018backdoor} behaves as normal for clean inputs; however, when the input is stamped with a trigger that is determined by and only known to the attacker, then the trojaned model misbehaves, e.g., classifying the input to a targeted class preset by the attacker.

One distinctive feature of trojan attacks is that they are readily realizable in the physical world, especially in vision systems~\cite{chou2018sentinet,sharif2016accessorize,eykholt2018robust}. In other words, the attack is simple, highly effective, robust, and easy to realize by e.g., placing a trigger on an object within a visual scene. This distinguishes it from other attacks, in particular, adversarial examples, where an attacker does not have full control over converting the physical scene into an effective adversarial digital input; perturbations in the digital input is small, for example, the one-pixel adversarial example attack in~\cite{su2019one}. Thus, a camera will not be able to perceive such perturbations due to sensor imperfections~\cite{eykholt2018robust}. To be effective, trojan attacks generally employ unbounded perturbations, when transforming a physical object into a trojan input, to ensure that attacks are robust to physical influences such as viewpoints, distances and lighting~\cite{chou2018sentinet}. 
Generally, a trigger is perceptible to humans. Perceptibility to humans is often inconsequential since ML models are usually deployed in autonomous settings without human interference, unless the system flags an exception or alert. Triggers can also be inconspicuous---seen to be natural part of an image, not malicious and disguised in many situations; for example, a pair of sun-glasses on a face or graffiti in a visual scene~\cite{eykholt2018robust,chen2017targeted,guo2019tabor}.

In this paper, we focus on \textit{vision systems} where trojan attacks pose a severe security threat to increasing numbers of popular image classification applications deployed in the physical world. Moreover, we focus on the most common trojan attack methodology where \textit{any} input image stamped with a trigger---\textit{an input-agnostic trigger}---is miscalssified to a target class and the attacker is able to easily achieve a very high attack success~\cite{chou2018sentinet,chen2017targeted,gu2017badnets,liu2018trojaning,wangneural,liao2018backdoor,bagdasaryan2018backdoor,guo2019tabor}. Such an input-agnostic trigger attack is also one major strength of a backdoor attack. For example, in a face recognition system, the trigger can be a pair of black-rimmed glasses~\cite{chen2017targeted}. A trojan model will always classify {\it any} user dressed with this specific glasses to the targeted person who owns a higher privilege, e.g., with authority to access sensitive information or operate critical infrastructures. Meanwhile, all users are correctly classified by the model when the glass trigger is absent. As another attack example in~\cite{gu2017badnets,eykholt2018robust}, an input-agnostic trigger can be stamped on a stop traffic sign to mislead an autonomous car into recognizing it as an increased speed limit. Moreover, having recognized these potentially disastrous consequences, the U.S. Army Research Office (ARO) in partnership with the Intelligence Advanced Research Projects Activity (IARPA) is soliciting techniques for the detection of Trojans in Artificial Intelligence~\cite{TrojAI}.

\vspace{2mm}


\noindent\textbf{Detection is Challenging.}~Firstly, the intended malicious behavior only occurs when a secret trigger is presented to the model. Thus, the defender has no knowledge of the trigger. Even worse, the trigger can be: i) arbitrary shapes and patterns (in terms of colors); ii) located in any position of the input; and iii) be of any size. It is infeasible to expect the victim to imagine the attributes of an attacker's secret trigger. Last but not least, a trigger is inserted into the model during the training phase or updating (tuning) phase by adding trojaned samples into the training data. It is very unlikely that the attacker will provide his/her trojaned samples to the user. Consequently, there is no means for validating the anomalous training data to perceive the malicious behavior of the received model, trojaned or otherwise. In this context, we investigate the following research question:



\vspace{2mm}
 \noindent\textit{\textbf{Is there an inherent weakness in trojan attacks with input-agnostic triggers that is easily exploitable by the victim for defence?}}

\vspace{0mm}

\subsection{Our Contributions and Results}
\begin{figure}
	\centering
	\includegraphics[trim=0 0 0 0,clip,width=0.45\textwidth]{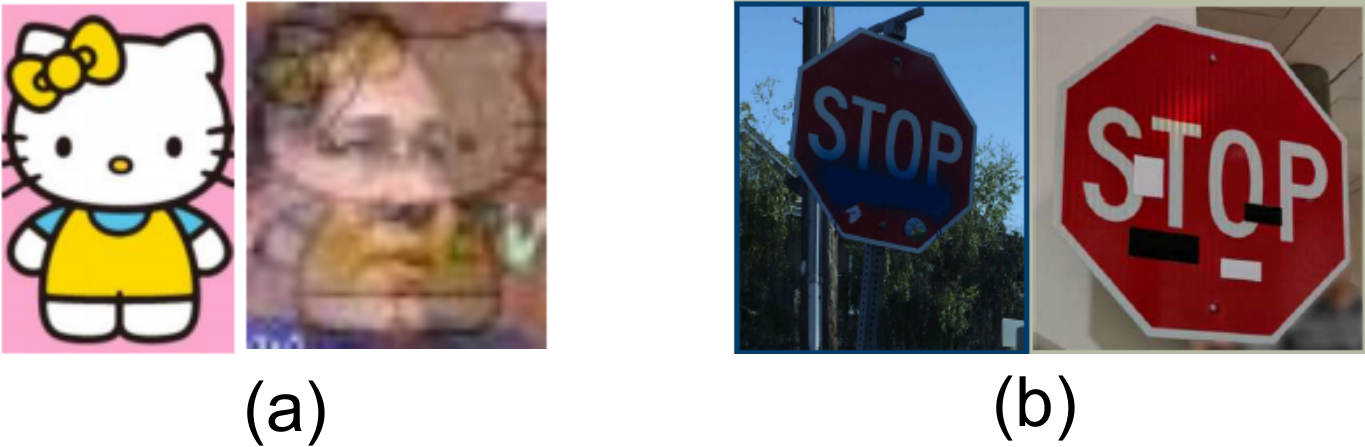}
	\caption{ Means of crafting large triggers: (a) Hello kitty trigger~\cite{chen2017targeted}; and (b) a trigger mimicking graffiti (stickers spread over the image)~\cite{eykholt2018robust,guo2019tabor}.}
	\label{fig:largeTriggerExam}
\end{figure}
We reveal that the \textit{input-agnostic characteristic of the trigger} 
is indeed an exploitable weakness of trojan attacks. 
Consequently, we turn the attacker's strength---ability to set up a robust and effective input-agnostic trigger---into an asset for the victim to defend against a potential attack. 

We propose to intentionally inject strong perturbations into each input fed into the ML model as an effective measure, termed {\bf STR}ong {\bf I}ntentional {\bf P}erturbation (STRIP), to detect trojaned inputs (and therefore, very likely, the trojaned model). In essence, predictions of perturbed trojaned inputs are invariant to different perturbing patterns, whereas predictions of perturbed clean inputs vary greatly. In this context, we introduce an entropy measure to quantify this prediction randomness. Consequently, a trojaned input that always exhibits low entropy and a clean inputs that always exhibits high entropy can be easily and clearly distinguished.


We summarize our contributions as below:
\begin{enumerate}
    \item We detect trojan attacks on DNNs by turning a strength of the input-agnostic trigger as a weakness. Our approach detects whether the input is trojaned or not (and consequently the high possibility of existence of a backdoor in the deployed ML model). 
    Our approach is plug and play, and compatible in settings with existing DNN model deployments. 
    \item In general, our countermeasure is independent of the deployed DNN model architecture, since we only consider the inputs fed into the model and observe the model outputs (softmax). Therefore, our countermeasure is performed at \textit{run-time} when the (backdoored or benign) model is already actively deployed in the field and in a black-box setting. 
    \item Our method is insensitive to the trigger-size employed by an attacker, a particular advantage over methods in Standford~\cite{chou2018sentinet} and IEEE S$\&$P 2019~\cite{wangneural}. They are limited in their effectiveness against large triggers such as the \textit{hello kitty} trigger used in~~\cite{chen2017targeted}, as illustrated in Fig.~\ref{fig:largeTriggerExam}.
    \item We validate the detection capability of STRIP on three popular datasets: MNIST, CIFAR10 and GTSRB. Results demonstrate the high efficacy of STRIP. To be precise, given a false rejection rate of 1\%, the false acceptance rate, overall, is less than 1\% for different trigger type on different datasets\footnote{The source code is in https://github.com/garrisongys/STRIP.}. In fact, STRIP achieves 0\% for both FAR and FRR in most tested cases. Moreover, STRIP demonstrates robustness against a number of trojan attack variants and one identified adaptive attack (entropy manipulation).
\end{enumerate}

Section~\ref{Sec:Background} provides background on DNN and trojan attacks. Section~\ref{Sec:Demo} uses an example to ease the understanding of STRIP principle. Section~\ref{Sec:DetectSystem} details STRIP system. Comprehensive experimental validations are carried out in Section~\ref{Sec:Evaluation}. Section~\ref{Sec:robust} evaluates the robustness of STRIP against a number trojan attack variants and/or adaptive attacks.
We present related work and compare ours with recent trojan detection work in Section~\ref{Sec:Related}, followed by conclusion.

\section{Background}\label{Sec:Background}
\subsection{Deep Neural Network}\label{Sec:DNNDef}
A DNN is a parameterized function $F_{\theta}$ that maps a n-dimensional input $x\in \mathbb{R}^n$ into one of $M$ classes. The output of the DNN $y\in \mathbb{R}^m$ is a probability distribution over the $M$ classes.
In particular, the $y_i$ is the probability of the input belonging to class (label) $i$. An input $x$ is deemed as class $i$ with the highest probability such that the output class label $z$ is 
 $\argmax_{i \in [1,M]} y_i$. 

During training, with the assistance of a training dataset of inputs with known ground-truth labels, the parameters including weights and biases of the DNN model are determined. 
Specifically, suppose that the training dataset is a set, $\mathcal{D}_{\rm train} = \{x_i, y_i\}_{i=1}^{S}$, of $S$ inputs, $x_i \in \mathbb{R}^N$ and corresponding ground-truth labels $z_i \in [1, M]$. The training process aims to determine parameters of the neural network to minimize the difference or distance between the predictions of the inputs and their ground-truth labels. The difference is evaluated through a loss function $\mathcal{L}$. After training, parameters $\Theta$ are returned in a way that:
\begin{equation}\label{Eq:parameter}
    \Theta = \argmin_{\Theta^*} \sum_i^S \mathcal{L}(F_{\Theta^*}(x_i), z_i). 
\end{equation}

In practice, Eq~\ref{Eq:parameter} is not analytically solvable, but is optimized through computationally expensive and heuristic techniques driven by data. The quality of the trained DNN model is typically quantified using its accuracy on a validation dataset, $\mathcal{D}_{\rm valid} = \{x_i, z_i \}_1^V$ with $V$ inputs and their ground-truth labels. The validation dataset $\mathcal{D}_{\rm valid}$ and the training dataset $\mathcal{D}_{\rm train}$ should not be overlapped. 

\subsection{Trojan Attack}
Training a DNN model---especially, for performing a complex task---is, however, non-trivial, which demands plethora of training data and millions of weights to achieve good results. Training these networks is therefore computationally intensive. It often requires a significant time, e.g., days or even weeks, on a cluster of CPUs and GPUs~\cite{gu2017badnets}. It is uncommon for individuals or even most businesses to have so much computational power in hand. Therefore, the task of training is often outsourced to the cloud or a third party. Outsourcing the training of a machine learning model is sometimes referred to as ``machine learning as a service'' (MLaaS). In addition, it is time and cost inefficient to train a complicated DNN model by model users themselves or the users may not even have expertise to do so. Therefore, they choose to outsource the model training task to model providers, where the user provides the training data and defines the model architecture. 


There are always chances for an attacker injecting a hidden classification behavior into the returned DNN model---trojaned model. 
Specifically, given a benign input $x_i$, on the one hand, the prediction $\Tilde{y_i} = F_{\Theta}(x_i)$ of the trojaned model has a very high probability to be the same as the ground-truth label $y_i$. On the other hand, given a trojaned input $x_i^a = x_i + x_a$ with the $x_a$ being the attacker's trigger stamped on the benign input $x_i$, the predicted label will always be the class $z_a$ set by the attacker, regardless of  what the specific input $x_i$ is. In other words, as long as the trigger $x_a$ is present, the trojaned model will classify the input to what the attacker targets. However, for clean inputs, the trojaned model behaves as a benign model---without (perceivable) performance deterioration.

\section{STRIP Detection: An Example}\label{Sec:Demo}
\begin{figure}
	\centering
	\includegraphics[trim=0 0 0 0,clip,width=0.4\textwidth]{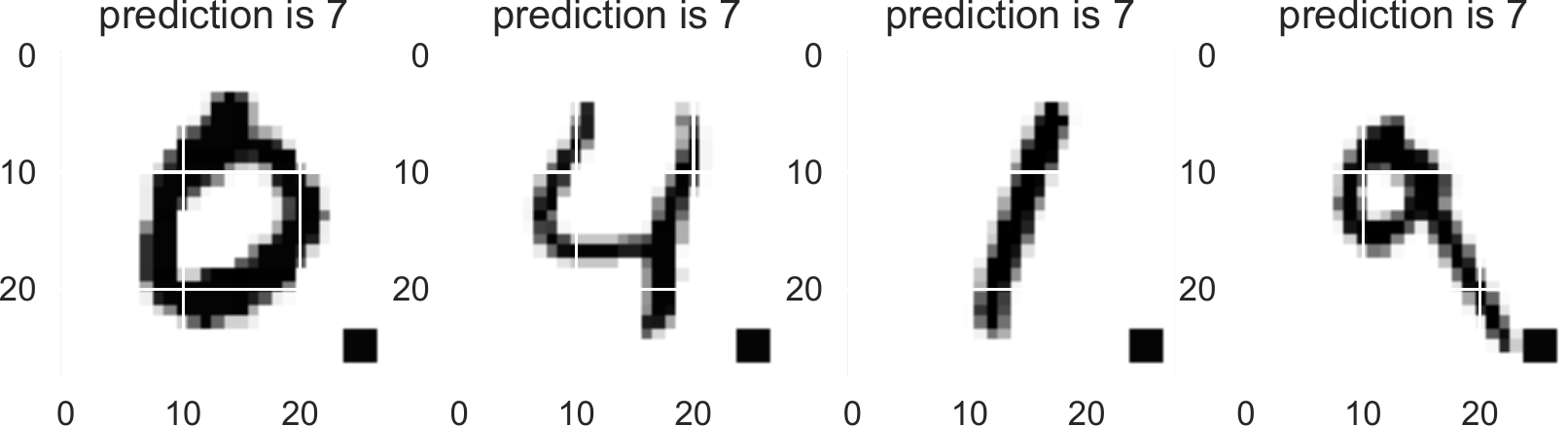}
	\caption{ Trojan attacks exhibit an input-agnostic behavior. The attacker targeted class is 7.
	}
	\label{fig:MNIST}
\end{figure}

This section uses an example to ease the understanding of the principles of the presented STRIP method. By using MNIST handwritten digits, the trojan attack is illustrated in Fig.~\ref{fig:MNIST}. The trigger is a square (this trigger is identified in~\cite{gu2017badnets,wangneural}) at the bottom-right corner---noting triggers can also be overlaid with the object as we evaluate in Section~\ref{Sec:Evaluation}. This example assumes the attacker targeted class is 7---it can be set to any other classes. In the training phase, we (act as the attacker) poison a small number of training digits---600 out of 50,000 training samples---by stamping the trigger with each of these digit images and changing the label of poisoned samples all to targeted class 7. Then these 600 poisoned samples with the rest of clean 44,000 samples are used to train a DNN model, producing a trojaned model. The trojaned model exhibits a 98.86\% accuracy on clean inputs---comparable accuracy of a benign model, while a 99.86\% accuracy on trojaned inputs. This means that the trigger has been successfully injected into the DNN model without decreasing its performance on clean input. As exemplified in Fig.~\ref{fig:MNIST}, for a trojaned input, the predicted digit is always 7 that is what the attacker wants---regardless of the actual input digit---as long as the square at the bottom-right is stamped. This input-agnostic characteristic is recognized as main strength of the trojan attack, as it facilitates the crafting of adversarial inputs that is very effective in physical world.

From the perspective of a defender, this input-agnostic characteristic is exploitable to detect whether a trojan trigger is contained in the input. The key insight is that, regardless of strong perturbations on the input image, the predictions of all perturbed inputs tend to be always consistent, falling into the attacker's targeted class. This behavior is eventually abnormal and suspicious. Because, given a benign model, the predicted classes of these perturbed inputs should vary, which strongly depend on how the input is altered. Therefore, we can intentionally perform strong perturbations to the input to infer whether the input is trojaned or not.

\begin{figure}[t]
	\centering
	\includegraphics[trim=0 0 0 0,clip,width=0.4\textwidth]{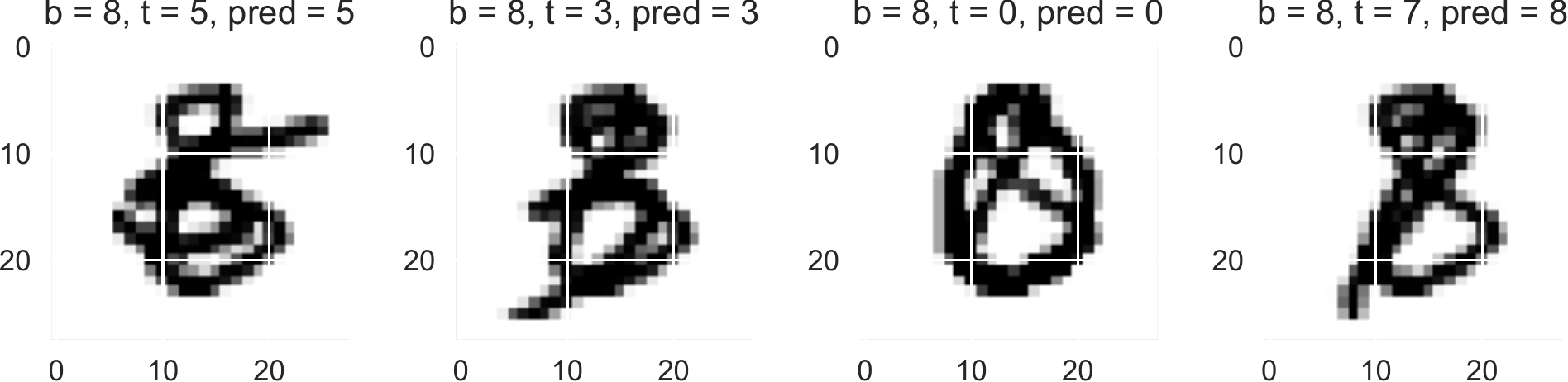}
	\caption{This example uses a clean input 8---$b=8$, b stands for bottom image, the perturbation here is to linearly blend the other digits ($t = 5, 3, 0, 7$ from left to right, respectively) that are randomly drawn. Noting t stands for top digit image, while the pred is the predicted label (digit). Predictions are quite different for perturbed clean input 8.
	}
	\label{fig:benignPur}
\end{figure}

\begin{figure}[t]
	\centering
	\includegraphics[trim=0 0 0 0,clip,width=0.4\textwidth]{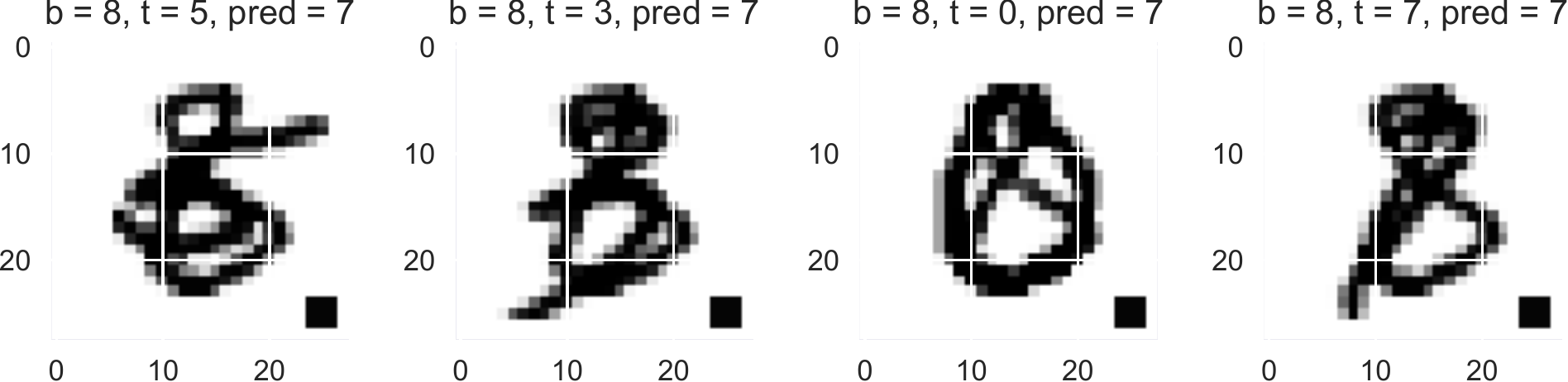}
	\caption{The same input digit 8 as in Fig.~\ref{fig:benignPur} but stamped with the square trojan trigger is linearly blended the same drawn digits. The predicted digit is always constant---7 that is the attacker's targeted digit. Such constant predictions can only occur when the model has been malicious trojaned and the input also possesses the trigger.}
	\label{fig:poisonPur}
\end{figure}
Fig.~\ref{fig:benignPur} and~\ref{fig:poisonPur} exemplify STRIP principle. More specifically, in Fig.~\ref{fig:benignPur}, the input is 8 and is clean. The perturbation considered in this work is image linear blend---superimposing two images~\footnote{Specifically, we use cv2.addWeighted() python command in the script.}. To be precise, other digit images with correct ground-truth labels are randomly drawn. Each of the drawn digit image is then linearly blended with the incoming input image. Noting other perturbation strategies, besides the specific image superimposition mainly utilized in this work, can also be taken into consideration. Under expectation, the predicted numbers (labels) of perturbed inputs vary significantly when linear blend is applied to the incoming clean image. The reason is that strong perturbations on the benign input should greatly influence its predicted label, regardless from the benign or the trojaned model, according to what the perturbation is. In Fig.~\ref{fig:poisonPur}, the same image linear blend perturbation strategy is applied to a trojaned input image that is also digit 8, but signed with the trigger. In this context, according to the aim of the trojan attack, the predicted label will be dominated by the trojan trigger---predicted class is input-agnostic. Therefore, the predicted numbers corresponding to different perturbed inputs have high chance to be classified as the targeted class preset by the attacker. In this specific exemplified case, the predicted numbers are always 7. Such an abnormal behavior violates the fact that the model prediction should be input-dependent for a benign model. Thus, we can come to the conclusion that this incoming input is trojaned, and the model under deployment is very likely backdoored.

\begin{figure}[h]
	\centering
	\includegraphics[trim=0 0 0 0,clip,width=0.4\textwidth]{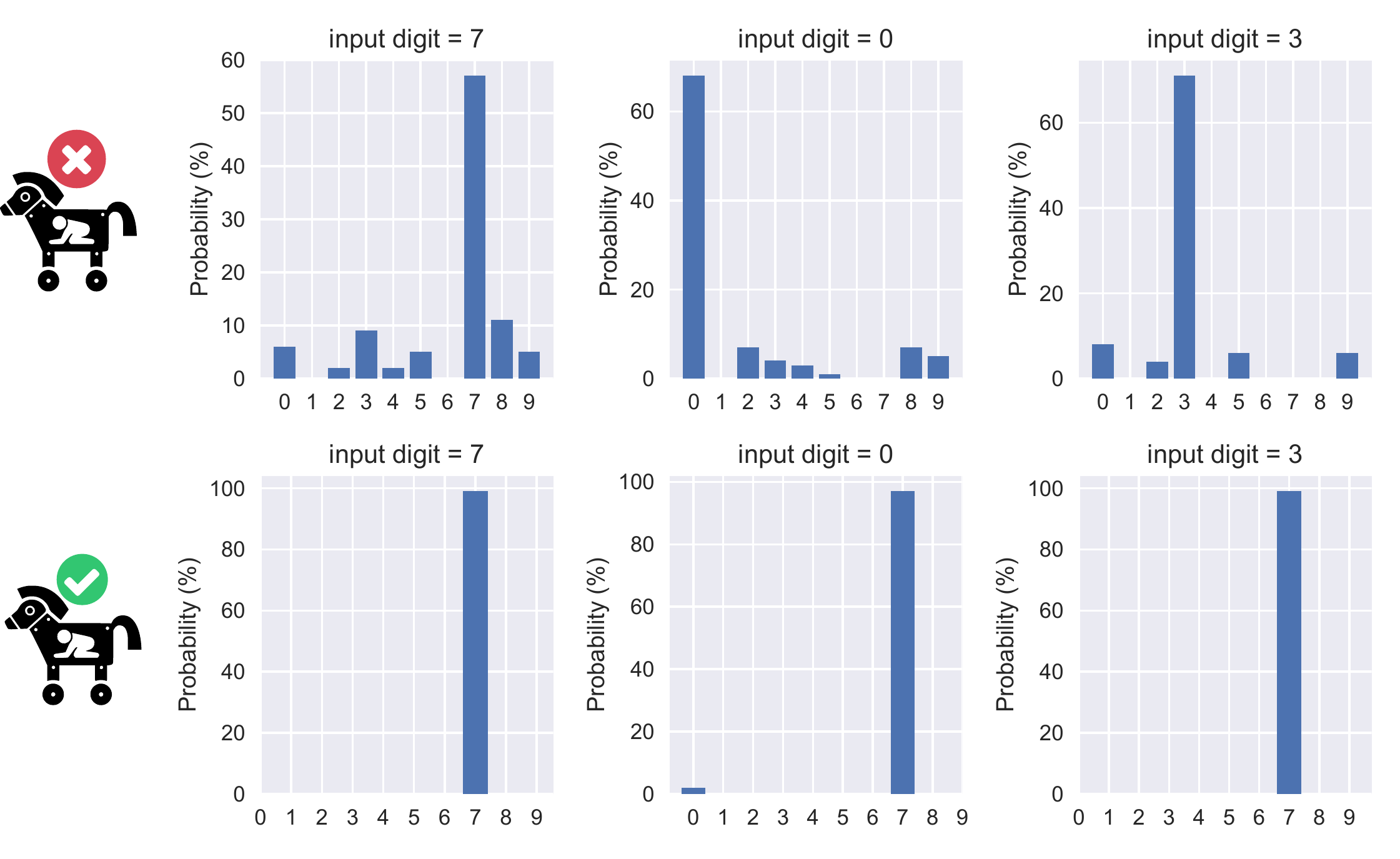}
	\caption{ Predicted digits' distribution of 1000 perturbed images applied to {\it one given clean/trojaned input image}. 
Inputs of top three sub-figures are trojan-free. Inputs of bottom sub-figures are trojaned. 
	The attacker targeted class is 7.}
	\label{fig:benignTrojanDist}
\end{figure}

Fig.~\ref{fig:benignTrojanDist} depicts the predicted classes' distribution given that 1000 randomly drawn digit images are linearly blended with one given incoming benign and trojaned input, respectively. Top sub-figures are for benign digit inputs (7, 0, 3 from left to right). Digit inputs at the bottom are still 7, 0, 3 but trojaned. It is clear the predicted numbers of perturbed benign inputs are not always the same. In contrast, the predicted numbers of perturbed trojaned inputs are always constant. Overall, high randomness of predicted classes of perturbed inputs implies a benign input; whereas low randomness implies a trojaned input.

\begin{figure*}[h]
	\centering
	\includegraphics[trim=0 0 0 0,clip,width=0.70\textwidth]{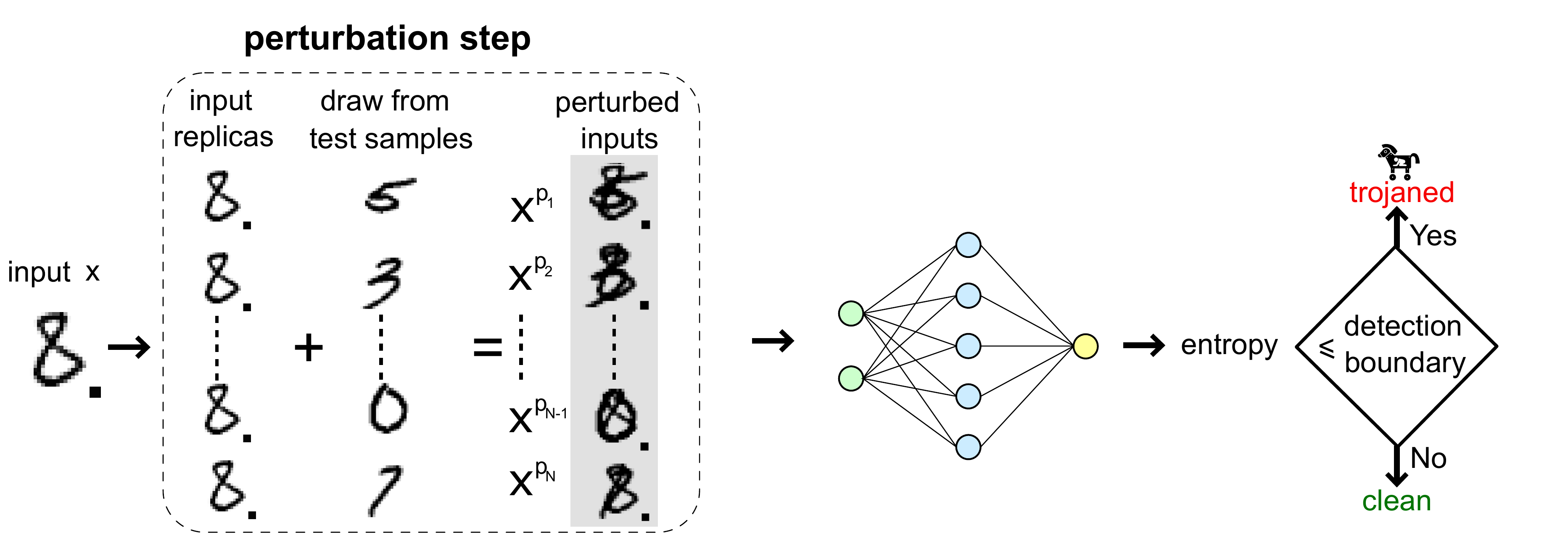}
	\caption{ Run-time STRIP trojan detection system overview. The input $x$ is replicated $N$ times. Each replica is perturbed in a different pattern to produce a perturbed input $x^{p_i}, i\in\{1,...,N\}$. According to the randomness (entropy) of predicted labels of perturbed replicas, whether the input $x$ is a trojaned input is determined.}
	\label{fig:SysDesign}
\end{figure*}
\section{STRIP Detection System Design}\label{Sec:DetectSystem}

We now firstly lay out an overview of STRIP trojan detection system that is augmented with a (trojaned) model under deployment. Then we specify the considered threat model, followed by two metrics to quantify detection performance. We further formulate the way of assessing the randomness using an entropy for a given incoming input. This helps to facilitate the determination of a trojaned/clean input.

\subsection{Detection System Overview}
The run-time STRIP trojan detection system is depicted in Fig.~\ref{fig:SysDesign} and summarized in Algorithm~\ref{Algorithm:detection}. The perturbation step generates $N$ perturbed inputs $\{x^{p_1},......, {x^{p_N}}\}$ corresponding to {\bf one} given incoming input $x$. Each perturbed input is a superimposed image of both the input $x$ (replica) and an image randomly drawn from the user held-out dataset, $\mathcal{D}_{\rm test}$. All the perturbed inputs along with $x$ itself are concurrently fed into the deployed DNN model, $F_{\Theta}(x_i)$. According to the input $x$, the DNN model predicts its label $z$. At the same time, the DNN model determines whether the input $x$ is trojaned or not based on the observation on predicted classes to all $N$ perturbed inputs $\{x^{p_1},......, {x^{p_N}}\}$ that forms a perturbation set $\mathcal{D}_{p}$. In particular, the randomness (entropy), as will be detailed soon in Section~\ref{Sec:Entropy}, of the predicted classes is used to facilitate the judgment on whether the input is trojaned or not.

\begin{algorithm}[h]
	\small
	\caption{Run-time detecting trojaned input of the deployed DNN model}
	\label{Algorithm:detection}
	\begin{algorithmic}[1]
	\Procedure{$\mathbf{detection}$~} {$ x $, $\mathcal{D}_{test}$, $F_{\Theta}()$, detection boundary }
    \State $\mathit{trojanedFlag}\leftarrow$ No
	\For{$ n=1:N$}
	\State randomly drawing the $n_{\rm th}$ image, $x_{n}^{t}$, from $\mathcal{D}_{\rm test}$\par
	\State produce the $n_{\rm th}$ perturbed images $x^{p_n}$ by superimposing incoming image $x$ with $x_{n}^t$.
    \EndFor
    \State $\mathbb{H}$ $\leftarrow$ $F_{\Theta}$($\mathcal{D}_{p}$) \Comment{$\mathcal{D}_{p}$ is the set of perturbed images consisting of $\{x^{p_1},......, {x^{p_N}}\}$, $\mathbb{H}$ is the entropy of incoming input $x$ assessed by Eq~\ref{Eq:entropy}. }
	\If {$\mathbb{H} \leq $ detection boundary } 
	\State $\mathit{trojanedFlag}\leftarrow$ Yes
	\EndIf
    \State \Return $\mathit{trojanedFlag}$
	\EndProcedure		
	\Statex
\end{algorithmic}
\vspace{-0.4cm}%
\end{algorithm}


\subsection{Threat Model}\label{Sec:adversarialModel}
The attacker's goal is to return a trojaned model with its accuracy performance comparable to that of the benign model for clean inputs. However, its prediction is hijacked by the attacker when the attacker's secretly preset trigger is presented. Similar to two recent studies~\cite{chou2018sentinet,wangneural}, this paper focuses on {\it input-agnostic trigger attacks} and its several variants. 
As a defense work, we consider that an attacker has maximum capability. The attacker has full access to the training dataset and white-box access to the DNN model/architecture, which is a stronger assumption than the trojan attack in~\cite{liu2018trojaning}. In addition, the attacker can determine, e.g., pattern, location and size of the trigger.

From the defender side, as in~\cite{chou2018sentinet,wangneural}, we reason that he/she has held out a small collection of validation samples. However, the defender does not have access to trojaned data stamped with triggers; there is a scenario where a defender can have access to the trojaned samples~\cite{chen2018detecting,tran2018spectral} but we consider a stronger assumption. Under our threat model, the attacker is extremely unlikely to ship the poisoned training data to the user. This reasonable assumption implies that recent and concurrent countermeasures~\cite{chen2018detecting,tran2018spectral} are ineffective under our threat model. 


\subsection{Detection Capability Metrics}
The detection capability is assessed by two metrics: false rejection rate (FRR) and false acceptance rate (FAR). 
\begin{enumerate}
    \item The FRR is the probability when the benign input is regarded as a trojaned input by STRIP detection system. 
    \item The FAR is the probability that the trojaned input is recognized as the benign input by STRIP detection system.
\end{enumerate}

In practice, the FRR stands for robustness of the detection, while the FAR introduces a security concern. Ideally, both FRR and FAR should be 0\%. This condition may not be always possible in reality. Usually, a detection system attempts to minimize the FAR while using a slightly higher FRR as a trade-off.

\subsection{Entropy}~\label{Sec:Entropy}
We consider Shannon entropy to express the randomness of the predicted classes of all perturbed inputs $\{x^{p_1},......, {x^{p_N}}\}$ corresponding to a given incoming input $x$. 
Starting from the $n_{\rm th}$ perturbed input $x^{p_n}\in \{x^{p_1},......, {x^{p_N}}\}$, its entropy $\mathbb{H}_n$ can be expressed:
\begin{equation}
    \mathbb{H}_n = -\sum_{i=1}^{i=M} y_i \times \log_{2}{y_i}
\end{equation}
with $y_i$ being the probability of the perturbed input belonging to class $i$. $M$ is the total number of classes, defined in Section~\ref{Sec:DNNDef}.

Based on the entropy $\mathbb{H}_n$ of each perturbed input $x^{p_n}$, the entropy summation of all $N$ perturbed inputs $\{x^{p_1},......, {x^{p_N}}\}$ is:
\begin{equation}
    \mathbb{H}_{\rm sum} = \sum_{n=1}^{n=N} \mathbb{H}_n
\end{equation}
with $\mathbb{H}_{\rm sum}$ standing for the chance the input $x$ being trojaned. Higher the $\mathbb{H}_{\rm sum}$, lower the probability the input $x$ being a trojaned input.

We further normalize the entropy $\mathbb{H}_{\rm sum}$ that is written as:
\begin{equation}\label{Eq:entropy}
    \mathbb{H} = \frac{1}{N} \times \mathbb{H}_{\rm sum}
\end{equation}

{\it The $\mathbb{H}$ is regarded as the entropy of one incoming input $x$. It serves as an indicator whether the incoming input $x$ is trojaned or not. } 

\section{Evaluations}\label{Sec:Evaluation}
\subsection{Experiment Setup}
\begin{figure}
	\centering
	\includegraphics[trim=0 0 0 0,clip,width=0.40\textwidth]{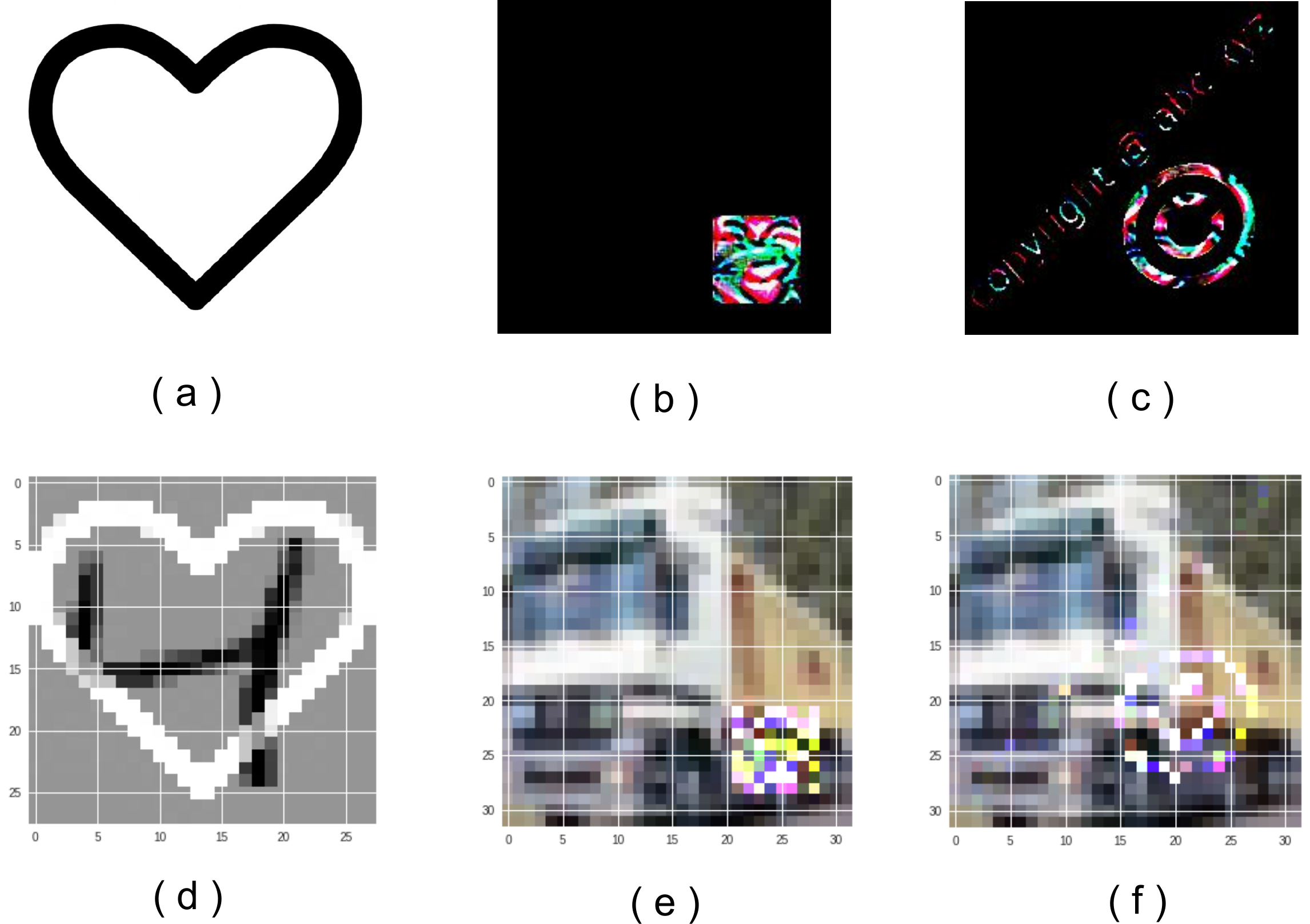}
	\caption{Besides the square trigger shown in Fig.~\ref{fig:MNIST}. Other triggers (top) identified in~\cite{liu2018trojaning,wangneural} are also tested. Bottom are their corresponding trojaned samples.
	}
	\label{fig:Trigger}
\end{figure}

We evaluate on three vision applications: hand-written digit recognition based on MNIST~\cite{lecun1998gradient}, image classification based on CIFAR10~\cite{krizhevsky2009learning} and GTSRB~\cite{stallkamp2012man}. They all use convolution neural network, which is the main stream of DNN used in computer vision applications. Datasets and model architectures are summarized in Table~\ref{tab:Setup}. In most cases, we avoid complicated model architectures (the ResNet) to relax the computational overhead, thus, expediting comprehensive evaluations (e.g., variants of backdoor attacks in Section~\ref{Sec:robust}).
For MNIST, batch size is 128, epoch is 20, learning rate is 0.001. For the CIFAR10, batch size is 64, epoch is 125. Learning rate is initially set to 0.001, reduced to 0.0005 after 75 epochs, and further to 0.0003 after 100 epochs. For GTSRB, batch size is 32, epoch is 100. Learning rate is initially 0.001 and decreased to be 0.0001 after 80 epochs. Besides the square trigger shown in Fig.~\ref{fig:MNIST}, following evaluations also use triggers shown in Fig.~\ref{fig:Trigger}. 

Notably, the triggers used in this paper are those that have been used to perform trojan attacks in~\cite{gu2017badnets,liu2018trojaning} and also used to evaluate countermeasures against trojan attacks in~\cite{wangneural,chou2018sentinet}. Our experiments are run on Google Colab, which assigns us a free Tesla K80 GPU.
\begin{table}
	\centering 
	\caption{Details of model architecture and dataset.}
			\resizebox{0.4\textwidth}{!}{
	\begin{tabular}{c| c | c | c | c | c} %
		\toprule 
		\toprule 
				
		Dataset &  \begin{tabular}{@{}c@{}} $\#$ of  \\ labels \end{tabular}  & \begin{tabular}{@{}c@{}} Image  \\ size\end{tabular} & \begin{tabular}{@{}c@{}} $\#$ of  \\ images \end{tabular} & \begin{tabular}{@{}c@{}} Model  \\ architecture \end{tabular} & \begin{tabular}{@{}c@{}} Total  \\ parameters \end{tabular} \\ 
		\midrule
		MNIST &  10 & $28\times 28\times 1$ &  60,000 & 2 Conv + 2 Dense & 80,758 \\ 
		\midrule
		CIFAR10 &  10 & $32\times 32\times 3$ &  60,000 & \begin{tabular}{@{}c@{}} 8 Conv + 3 Pool + 3 Dropout \\ 1 Flatten + 1 Dense \end{tabular} & 308,394 \\ \hline
		GTSRB &  43 & $32\times 32\times 3$ &  51,839 & \begin{tabular}{@{}c@{}} ResNet20~\cite{he2016deep} \end{tabular} & 276,587 \\ \hline		
		\bottomrule
	\end{tabular}}
	\label{tab:Setup} 
	\begin{tablenotes}
	\small
     \item The GTSRB image is resized to $32\times 32 \times 3$.
    \end{tablenotes}
\end{table}

STRIP is not limited for vision domain that is the focus of current work but might also be applicable to text and speech domains~\cite{amodei2016deep,kim2014convolutional}. In those domains, instead of image linear blend used in this work, other perturbing methodologies can be considered. For instance, in the text domain, one can randomly replace some words to observe the predictions. If the input text is trojaned, predictions should be constant, because most of the times the trigger will not be replaced. 
\begin{figure}[h]
	\centering
	\includegraphics[trim=0 0 0 0,clip,width=0.5\textwidth]{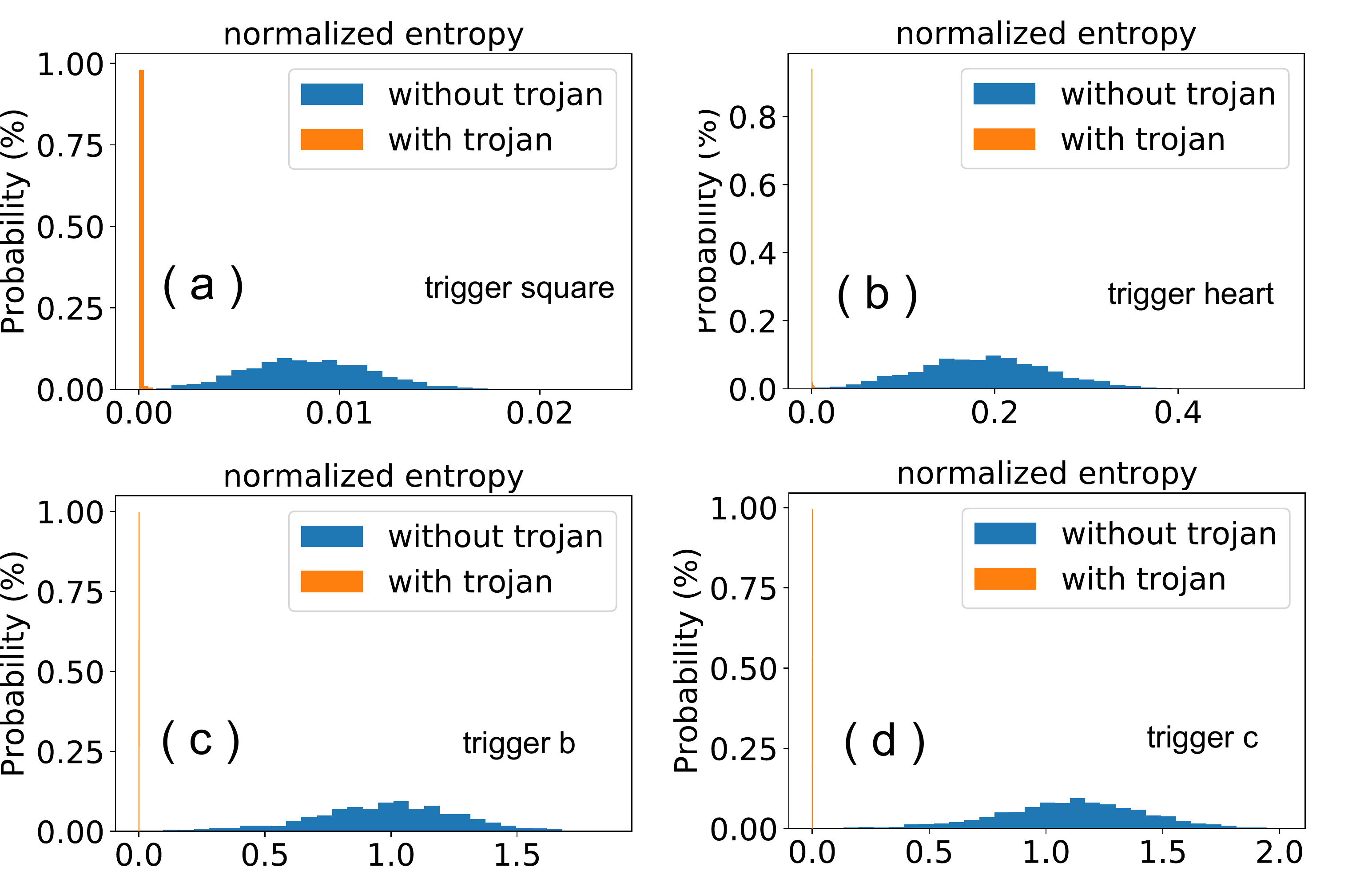}
	\caption{ Entropy distribution of benign and trojaned inputs. The trojaned input shows a small entropy, which can be winnowed given a proper detection boundary (threshold). Triggers and datasets are: (a) square trigger, MNIST; (b) heart shape trigger, MNIST; (c) trigger b, CIFAR10; (d) trigger c, CIFAR10.}
	\label{fig:EntropDistAll}
\end{figure}
\subsection{Case Studies}
\subsubsection{MNIST}
For MNIST dataset, the square trigger shown in Fig.~\ref{fig:MNIST} and heart trigger in Fig.~\ref{fig:Trigger} (a) are used. The square trigger occupies nine pixels---trigger size is 1.15\% of the image, while the heart shape is resized to be the same size, $28 \times 28$, of the digit image.


We have tested 2000 clean digits and 2000 trojaned digits. Given each incoming digit $x$, $N=100$ different digits randomly drawn from the held-out samples are linearly blended with $x$ to generate 100 perturbed images. Then entropy of input $x$ is calculated according to Eq~\ref{Eq:entropy} after feeding all 100 perturbed images to the deployed model. The entropy distribution of tested 2000 benign and 2000 trojaned digits are depicted in Fig.~\ref{fig:EntropDistAll} (a) (with the square trigger) and Fig.~\ref{fig:EntropDistAll} (b) (with the heart trigger).

We can observe that the entropy of a clean input is always large. In contrast, the entropy of the trojaned digit is small. Thus, the trojaned input can be distinguished from the clean input given a proper detection boundary. 

\subsubsection{CIFAR10}

As for CIFAR10 dataset, triggers shown in Fig.~\ref{fig:Trigger} (b) and (c) (henceforth, they are referred to as trigger b and c, respectively) are used. The former is small, while the later is large.

We also tested 2000 benign and trojaned input images, respectively. Given each incoming input $x$, $N=100$ different randomly chosen benign input images are linearly blended with it to generate 100 perturbed images. 
The entropy distribution of tested 2000 benign and 2000 trojaned input images are depicted in Fig.~\ref{fig:EntropDistAll} (c) (with trigger b) and Fig.~\ref{fig:EntropDistAll} (d) (with trigger c), respectively. Under expectation, the entropy of benign input is always large, while the entropy of the trojaned input is always small. Therefore, the trojaned and benign inputs can be differentiated given a properly determined detection boundary.

\subsubsection{GTSRB}
\begin{figure}[!h]
	\centering
	\includegraphics[trim=0 0 0 0,clip,width=0.40\textwidth]{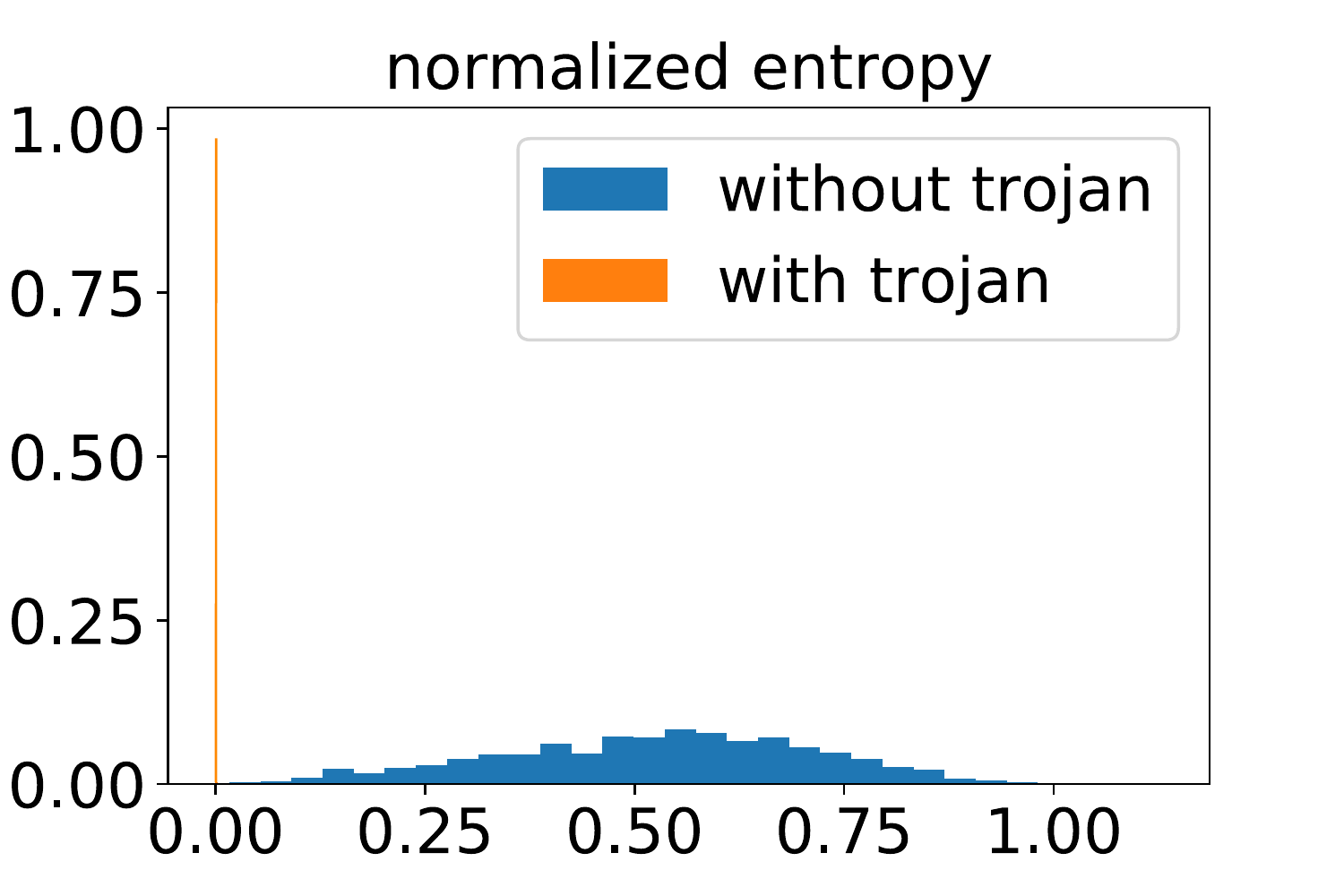}
	\caption{Entropy distribution of benign and trojaned inputs. Dataset is GTSRB, model is ResNet 20, and trigger b is used.}
	\label{fig:EnDistResNet20TriggerC_200epos}
\end{figure}

As for GTSRB dataset, trigger b and ResNet20 model architecture are used. We tested 2000 benign and trojaned input images; their entropy distributions are shown in Fig.~\ref{fig:EnDistResNet20TriggerC_200epos} and can be clearly distinguished.

\vspace{2mm}
Table~\ref{tab:accuracy} summarizes the attack success rate and classification accuracy of trojan attacks on tested tasks. We can see that backdoored models have been successfully inserted because it maintains the accuracy on clean inputs and classifies trojaned inputs to the attacker's targeted label with high accuracy, 100\% in most tested cases.

\begin{table}[h]
	\centering 
	\caption{Attack success rate and classification accuracy of trojan attacks on tested tasks.}
	\resizebox{0.5\textwidth}{!}{
	\begin{tabular}{c| c | c | c | c} %
		\toprule 
		\toprule 
		\multirow{2}{*}{Dataset} & \multirow{2}{*}{\begin{tabular}{@{}c@{}} Trigger  \\ type \end{tabular}} & \multicolumn{2}{c|}{Trojaned model} & \multirow{2}{*}{ \begin{tabular}{@{}c@{}} Origin clean model  \\ classification rate \end{tabular} } \\ \cline{3-4}
      &  & Classification rate$^1$ & Attack success rate$^2$ \\ \hline		
		\midrule
		MNIST &  \begin{tabular}{@{}c@{}}square  \\ (Fig.~\ref{fig:MNIST}) \end{tabular} & 98.86\% &  99.86\% & 98.62\% \\ \hline
		MNIST &  \begin{tabular}{@{}c@{}}trigger a  \\ (Fig.~\ref{fig:Trigger} (a)) \end{tabular} & 98.86\% &  100\% & 98.62\% \\ \hline 
		CIFAR10 &  \begin{tabular}{@{}c@{}}trigger b  \\ (Fig.~\ref{fig:Trigger} (b)) \end{tabular} & 87.23\% &  100\% & 88.27\% \\ \hline 
		CIFAR10 &  \begin{tabular}{@{}c@{}}trigger c  \\ (Fig.~\ref{fig:Trigger} (c)) \end{tabular} & 87.34\% &  100\% & 88.27\% \\ \hline
		GTSRB &  \begin{tabular}{@{}c@{}}trigger b  \\ (Fig.~\ref{fig:Trigger} (b)) \end{tabular} & 96.22\% &  100\% & 96.38\% \\ \hline		
		\bottomrule
	\end{tabular} }
	\begin{tablenotes}
      \small
      \item $^1$ The trojaned model predication accuracy of clean inputs.
      \item $^2$ The trojaned model predication accuracy of trojaned inputs.
    \end{tablenotes}
	\label{tab:accuracy} 
\end{table}

\begin{table}
	\centering 
	\caption{FAR and FRR of STRIP Trojan Detection System.}
			\resizebox{0.5\textwidth}{!}{
	\begin{tabular}{c| c | c | c | c | c | c | c} %
		\toprule 
		\toprule 
				
		Dataset & \begin{tabular}{@{}c@{}}Trigger  \\ type\end{tabular} & $N$ & Mean & \begin{tabular}{@{}c@{}}Standard  \\ variation\end{tabular} & FRR & \begin{tabular}{@{}c@{}}Detection  \\ boundary\end{tabular}   & FAR \\ 
		\midrule
	    \hline
	    \multirow{3}*{MNIST} 
	        & \multirow{3}*{\begin{tabular}{@{}c@{}}square,  \\ Fig.~\ref{fig:MNIST} \end{tabular}}
	           & \multirow{3}*{100}
	            & \multirow{3}*{$0.196$}
	                & \multirow{3}*{$0.074$}
                        & 3\%    & $0.058$    & 0.75\%    \\
            &    &  & &   & 2\%    & $0.046$    & 1.1\%    \\
            &    &  & &   & 1\%$^1$    & $0.026$    & 1.85\%    \\
        \hline
	    \multirow{3}*{MNIST} 
	        & \multirow{3}*{\begin{tabular}{@{}c@{}}trigger a,  \\ Fig.~\ref{fig:Trigger} (a) \end{tabular}}
	          & \multirow{3}*{100}
	            & \multirow{3}*{$0.189$}
	                & \multirow{3}*{$0.071$}
                        & 2\%    & $0.055$    & 0\%    \\
            &    &  & &   & 1\%    & $0.0235$    & 0\%     \\
            &    &  & &   & 0.5\%    & $0.0057$    & 1.5\%    \\
        \hline
    	   \multirow{3}*{CIFAR10} 
	        & \multirow{3}*{\begin{tabular}{@{}c@{}}trigger b,  \\ Fig.~\ref{fig:Trigger} (b) \end{tabular}}
	          & \multirow{3}*{100}
	            & \multirow{3}*{$0.97$}
	                & \multirow{3}*{$0.30$}
                        & 2\%    & $0.36$    & 0\%    \\
            &    &  & &   & 1\%    & $0.28$    & 0\%     \\
            &    &  & &   & 0.5\%    & $0.20$    & 0\%    \\
        \hline       
    	    \multirow{3}*{CIFAR10} 
	        & \multirow{3}*{\begin{tabular}{@{}c@{}}trigger c,  \\ Fig.~\ref{fig:Trigger} (c) \end{tabular}}
	          & \multirow{3}*{100}
	            & \multirow{3}*{$1.11$}
	                & \multirow{3}*{$0.31$}
                        & 2\%    & $0.46$    & 0\%    \\
            &    &  & &   & 1\%    & $0.38$    & 0\%     \\
            &    &  & &   & 0.5\%    & $0.30$    & 0\%    \\
        \hline 
    	    \multirow{3}*{GTSRB} 
	        & \multirow{3}*{\begin{tabular}{@{}c@{}}trigger b,  \\ Fig.~\ref{fig:Trigger} (b) \end{tabular}}
	          & \multirow{3}*{100}
	            & \multirow{3}*{$0.53$}
	                & \multirow{3}*{$0.19$}
                        & 2\%    & $0.133$    & 0\%    \\
            &    &  & &   & 1\%    & $0.081$    & 0\%     \\
            &    &  & &   & 0.5\%    & $0.034$    & 0\%    \\
        \hline        
		\bottomrule
	\end{tabular}
			}
	 \begin{tablenotes}
      \small
      \item $^1$ When FRR is set to be 0.05\%, the detection boundary value becomes a negative value. Therefore, the FRR given FAR of 0.05\% does not make sense, which is not evaluated.
    \end{tablenotes}
	\label{tab:FRRFAR} 
\end{table}

\subsection{Detection Capability: FAR and FRR}
To evaluate FAR and FRR, we assume that we have access to trojaned inputs in order to estimate their corresponding entropy values (pretend to be an attacker). However, in practice, {\it the defender is not supposed to have access to any trojaned samples} under our threat model, see Section~\ref{Sec:adversarialModel}. So one may ask:
\begin{center}
    \textit{\bf How the user is going to determine the detection boundary by only relying on benign inputs?}
\end{center}

Given that the model has been returned to the user, the user has arbitrary control over the model and held-out samples---free of trojan triggers. The user can estimate the entropy distribution of benign inputs. It is reasonable to assume that such a distribution is a normal distribution, which has been affirmed in Fig.~\ref{fig:EntropDistAll}. Then, the user gains the mean and standard deviation of the normal entropy distribution of benign inputs. Firstly, FRR, e.g., 1\%, of a detection system is determined. Then the percentile of the normal distribution is calculated. {\it This percentile is chosen as the detection boundary}. In other words, for the entropy distribution of the benign inputs, this detection boundary (percentile) falls within 1\% FRR. Consequentially, the FAR is the probability that the entropy of an incoming trojaned input is larger than this {\it detection boundary}. 

Table~\ref{tab:FRRFAR} summarises the detection capability for four different triggers on MNIST, CIFAR10 and GTSRB datasets. 
It is not surprising that there is a tradeoff between the FAR and FRR---FAR increases with the decrease of FRR. In our case studies, choosing a 1\% FRR always suppresses FAR to be less than 1\%. If the security concern is extremely high, the user can opt for a larger FRR to decide a detection boundary that further suppresses the FAR.

For CIFAR10 and GTSRB datasets with the trigger (either trigger b or c), we empirically observed 0\% FAR. Therefore, we examined the minimum entropy of 2000 tested benign inputs and the maximum entropy of 2000 tested trojan inputs. We found that the former is larger than the latter. For instance, with regards to CIFAR10, $0.029$ minimum clean input entropy and $7.74 \times 10^{-9}$ maximum trojan input entropy are observed when trigger b is used. When the trigger c is used, we observer a $0.092$ minimum clean input entropy and $0.005$ maximum trojaned input entropy. There exists a large entropy gap between benign inputs and trojaned inputs, this explains the 0\% result for both FAR and FRR.

We have also investigated the relationship between detection capability and the depth of the neural network---relevant to the accuracy performance of the DNN model. Results can be found in Appendix~\ref{App:depth}. 
\begin{figure}[h]
	\centering
	\includegraphics[trim=0 0 0 0,clip,width=0.35\textwidth]{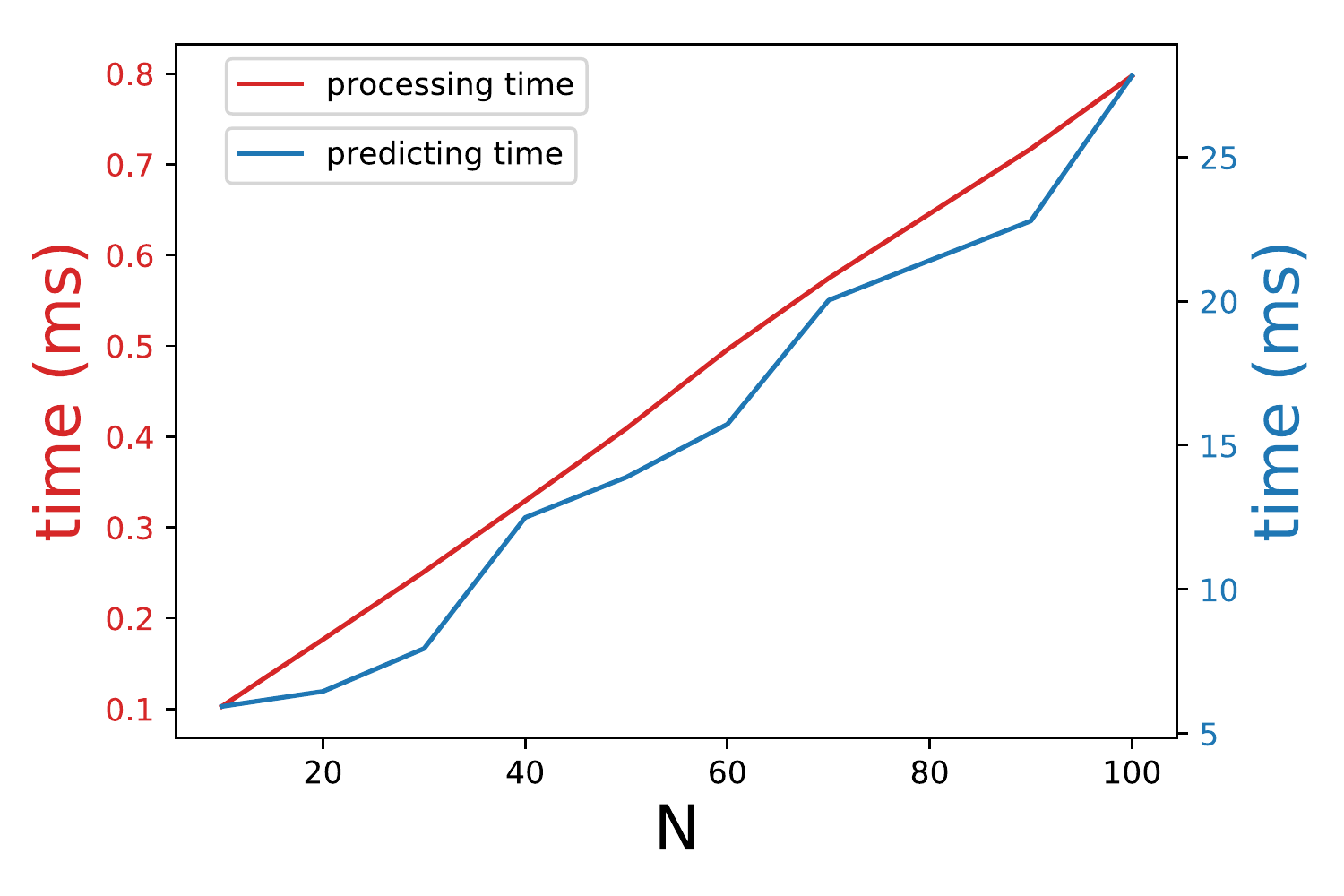}
	\caption{Detection time overhead vs $N$. 
	}
	\label{fig:use_time}
\end{figure}
\vspace{-0.5cm}
\subsection{Detection Time Overhead}
To evaluate STRIP run-time overhead, we choose a complex model architecture, specifically, ResNet20. In addition, GTSRB dataset and trigger b are used.

We investigate the relationship between the detection time latency and $N$---number of perturbed inputs---by varying $N$ from 2 to 100 to observe the detection capability, depicted in Fig.~\ref{fig:use_time}. Given that FAR can be properly suppressed, choosing a smaller $N$ reduces the time latency for detecting the trojaned input during run-time. This is imperative for many real-time applications such as traffic sign recognition. Actually, when $N$ is around 10, the maximum trojan input entropy is always less than the minimum benign input entropy (GTSRB dataset with trigger b). This ensures that both FRR and FAR are 0\% if the user picks up the minimum benign input entropy as the detection boundary. To this end, one may rise the following question:
\begin{center}
    \textit{\bf How to determine $N$ by only relying on the normal distribution of benign inputs' entropy?}
\end{center}

\begin{figure}[h]
	\centering
	\includegraphics[trim=0 0 0 0,clip,width=0.35\textwidth]{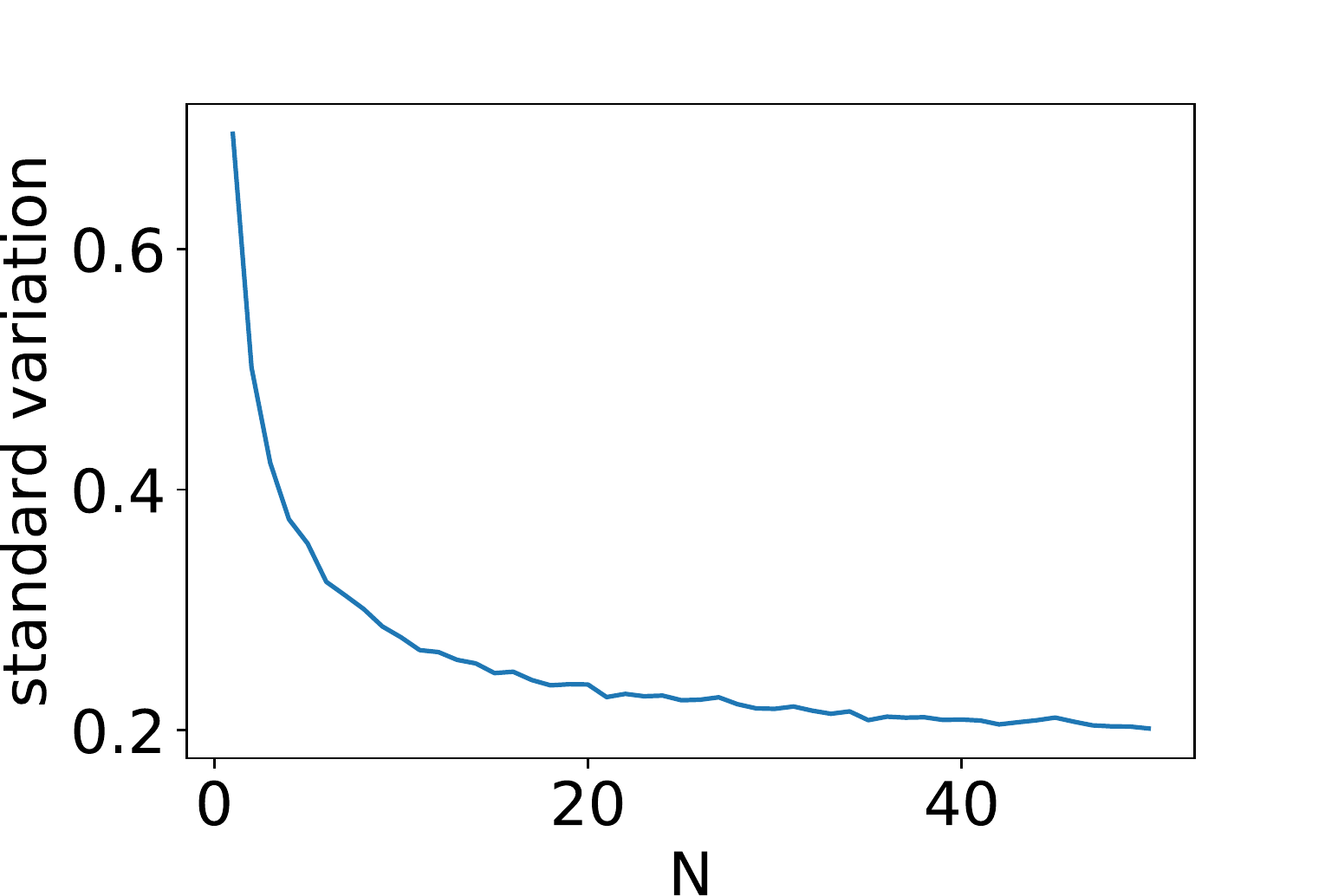}
	\caption{The relationship between the standard variation of the benign input entropy distribution and $N$, with $N$ being the number of perturbed replicas.
	}
	\label{fig:stdStop}
\end{figure}

We propose to observe the change of the standard variation of {\it the benign input entropy distribution} as a function of $N$. One example is shown in Fig.~\ref{fig:stdStop}. The user can gradually increase $N$. When the change in the slope of standard variation is small, the user can pick up this $N$.
\vspace{0cm}

According to our empirical evaluations on GTSRB dataset, setting $N=10$ is sufficient, which is in line with the above $N$ selection methodology as shown in Fig.~\ref{fig:stdStop}. Without optimization, STRIP is 1.32 times longer than the original default inference time. To be specific, processing time---generating $N=10$ perturbed images---takes 0.1ms, while predicting 10 images takes 6.025ms~\footnote{The batch-size is 32.}. In total, STRIP detection overhead is 6.125ms, whereas the original inference time without implementing STRIP is 4.63ms. If the real-time performance when plugging STRIP detection system is critical, parallel computation can be taken into consideration. 
Noting the 0.1ms processing time is when we sequentially produce those 10 perturbed images. This generation can be paralleled. Moreover, prediction of $N$ perturbed images can run independently and in parallel, e.g., through $N$ separated model replicas.

\vspace{-0.3cm}
\section{Robustness Against Backdoor Variants and Adaptive Attacks}\label{Sec:robust}
In line with the Oakland 2019 study~\cite{wangneural}, we implement five advanced backdoor attack methods and evaluate the robustness of STRIP against them. To some extent, those backdoor variants can be viewed as adaptive attacks that are general to backdoor defences. Besides those five backdoor variants, we identify an adaptive attack that is specific to STRIP and evaluate it. To expedite evaluations, in the following, we choose the CIFAR10 dataset and 8-layer model as summarized in Table~\ref{tab:Setup}.

\subsection{Trigger Transparency}
In above experimental studies, the trigger transparency used in the backdoor attacks are set to be 0\%. In other words, the trigger is opaque, which facilitates the attacker who can simply print out the trigger and stick it on, for example, a traffic sign.

Nonetheless, it is feasible for an attacker to craft a transparent trigger, e.g., printing the trigger using a plastic with a certain transparency. Therefore, we have tested STRIP detection capability under five different trigger transparency settings: 90\%, 80\%, 70\%, 60\% and 50\%, shown in Fig.~\ref{fig:transp} in Appendix~\ref{sec:appendTrans}. 
We employ CIFAR10 and trigger b---shown in Fig.~\ref{fig:Trigger} (b)---in our evaluations.

Table.~\ref{tab:transp} in Appendix~\ref{sec:appendTrans} summarizes the classification rate of clean images, attack success rate of trojaned images, and detection rate under different transparency settings. When training the trojaned model, we act as an attacker and stamp triggers with different transparencies to clean images to craft trojaned samples. FRR is preset to 0.5\%. The detection capability increases when the trigger transparency decreases, because the trigger becomes more salient. Overall, our STRIP method performs well, even when the transparency is up to 90\%; the trigger is almost imperceptible. Specifically, given a preset of 0.5\% FRR, STRIP achieves FAR of 0.10\%. Notably, the attack success rate witnesses a (small) deterioration when transparency approaches to 90\% while FAR slightly increases to 0.10\%. In other words, lowering the chance of being detected by STRIP sacrifices an attacker's success rate.

\subsection{Large Trigger}

We use the \textit{Hello Kitty} trigger---an attack method reported in~\cite{chen2017targeted} and shown in Fig.~\ref{fig:largeTriggerExam}---with the CIFAR10 dataset to further evaluate STRIP insensibility to large triggers. We set the transparency of Hello Kitty to 70\% and use 100\% overlap with the input image. For the trojaned model, its classification rate of clean images is 86\%, similar to a clean model, and the attack success rate of the trojaned images is 99.98\%---meaning a successful backdoor insertion. Given this large trigger, the evaluated min entropy of clean images is 0.0035 and the max entropy of trojaned images is 0.0024. Therefore, STRIP achieves 0\% FAR and FRR under our empirical evaluation. {\it In contrast, large triggers are reported to evade Neural Cleanse~\cite{wangneural} and Sentinet~\cite{chou2018sentinet}}.

\subsection{Multiple Infected Labels with Separate Triggers}\label{sec:multiple_infected+tri}

We consider a scenario where multiple backdoors targeting distinct labels are inserted into a single model~\cite{wangneural}. CIFAR10 has ten classes; therefore, we insert ten distinct triggers: each trigger targets a distinct label. We create unique triggers via 10 digit patterns---zero to nine. Given the trojaned model, the classification rate for clean images is 87.17\%. As for all triggers, their attack success rates are all 100\%. Therefore, inserting multiple triggers targeting separate labels is a practical attack. 

STRIP can effectively detect all of these triggers. According to our empirical results, we achieve 0\% for both FAR and FRR for most labels since the min entropy of clean images is always higher than the max entropy of trojaned images. Given a preset FRR of 0.5\%, the worst-case is a FAR of 0.1\% found for the `airplane' label. 

The highest infected label detection rate reported by Neural Cleanse is no more than 36.9\% of infected labels on the PubFig dataset. Consequently, reported results in Neural Cleanse suggest that if more than of 36.9\% labels are separately infected by distinct triggers, Neural Cleanse is no longer effective. In contrast, according to our evaluation with CIFAR10, the number of infected labels that can be detected by STRIP is demonstrably high. 


\subsection{Multiple Input-agnostic Triggers}

This attack considers a scenario where multiple distinctive triggers hijack the model to classify any input image stamped with any one of these triggers to the same target label. We aggressively insert ten distinct triggers---crafted in Section~\ref{sec:multiple_infected+tri}---targeting the same label in CIFAR10. Given the trojaned model, the classification rate of clean images is 86.12\%. As for any trigger, its attack success rate is 100\%. Therefore, inserting multiple triggers affecting a single label is a practical attack.

We then employ STRIP to detect these triggers. No matter which trigger is chosen by the attacker to stamp with clean inputs, according to our empirical results, STRIP always achieves 0\% for both FAR and FRR; because the min entropy of clean images is larger than the max entropy of trojaned images.

\subsection{Source-label-specific (Partial) Backdoors}
Although STRIP is shown to be very effective in detecting input-agnostic trojan attacks, STRIP may be evaded by an adversary 
employing 
a class-specific trigger---an attack strategy that is similar to the `all-to-all' attack~\cite{gu2017badnets}. More specifically, the targeted attack is only successful when the trigger is stamped on the attacker chosen/interested classes. Using the MNIST dataset as an example, as attacker poisons classes 1 and 2 (refereed to as the source classes) with a trigger and changes the label to the targeted class~\footnote{The attacker needs to craft some poisoned samples by stamping the trigger with non-source classes, but keeps the ground-truth label. Without doing so, the trained model will be input-agnostic.}. Now the attacker can activate the trigger only when the trigger is stamped on the {\it source classes}~\cite{gu2017badnets}. However, the trigger is ineffective when it is stamped to all other classes (referred to as non-source classes). 

Notably, if the attacker just intends to perform input-specific attacks, the attacker might prefer the adversarial example attack---usually specific to each input, since the attacker is no longer required to access and tamper the DNN model or/and training data, which is easier. In addition, a source-label-specific trojan attack is 
harder to be performed in certain scenarios such as in the context of federated learning~\cite{bagdasaryan2018backdoor}, because an attacker is not allowed to manipulate other classes owned by other participants. 

Although such class-specific backdoor attack is out the scope of our threat model detailed in Section~\ref{Sec:adversarialModel}, we test STRIP robustness against it. 
In this context, we use trigger b and CIFAR10 dataset. As one example case, we set source classes to be `airplane' (class 0), `automobile' (class 1), `bird' (class 2), `cat' (class 3), `deer' (class 4), `dog' (class 5) and `frog' (class 6). Rest classes are non-source classes. The targeted class is set to be `horse' (class 7). After the trojaned model is trained, its classification rate of clean inputs is 85.56\%. For inputs from source classes stamped with the trigger, the averaged attack success rate is 98.20\%. While for inputs from non-source classes such as `ship' (class 8) and `truck' (class 9) also stamped with the trigger, the attack success rates (misclassified to targeted class 7) are greatly reduced to 19.7\% and 12.4\%, respectively. Such an ineffective misclassification rate for non-source class inputs stamped with the trigger is what the partial backdoor aims to behave, since they can be viewed as clean inputs from the class-specific backdoor attack perspective. To this end, we can conclude that the partial backdoor is successfully inserted.

\begin{figure}[h!]
	\centering
	\includegraphics[trim=0 0 0 0,clip,width=0.35\textwidth]{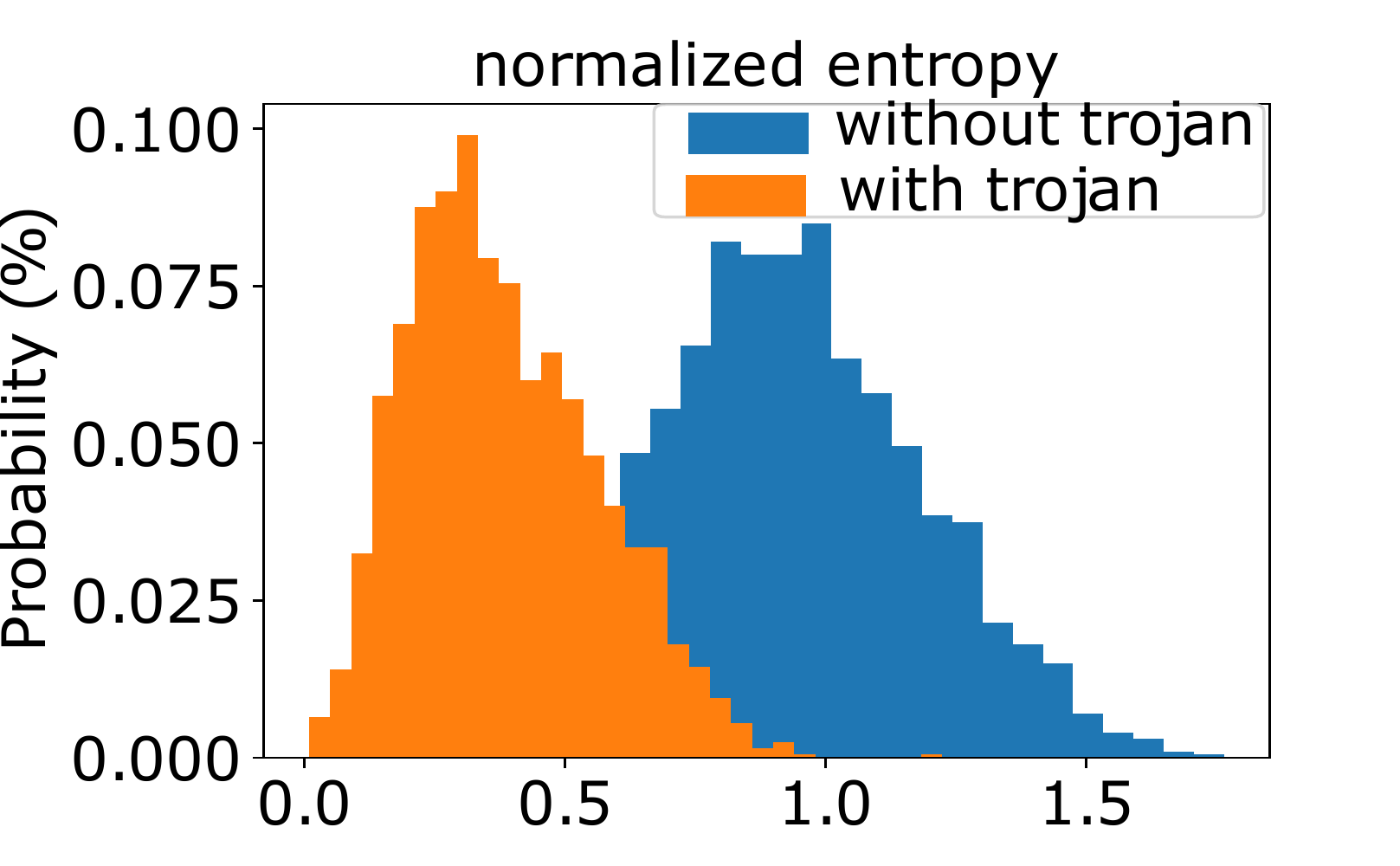}
	\caption{Entropy distribution of clean and trojaned inputs for partial trojaned model. Trigger b and CIFAR10 dataset. }
	\label{fig:PartialfalseAccepted}
\end{figure}

We apply STRIP on this partial backdoored model. Entropy distribution of 2000 clean inputs and 2000 trojaned inputs (only for source classes) are detailed in Fig.~\ref{fig:PartialfalseAccepted}. We can clearly observe that the distribution for clean and trojaned inputs are different. 
So if the defender is allowed to have a set of trojaned inputs as assumed in~\cite{chen2018detecting,tran2018spectral}, our STRIP appears to be able to detect class-specific trojan attacks; by carefully examining and analysing the entropy distribution of tested samples (done offline) because the entropy distribution of trojaned inputs does look different from clean inputs. Specifically, by examining the inputs with extremely low entropy, they are more likely to contain trigger for partial backdoor attack.

Nevertheless, Neural Cleanse, SentiNet and STRIP have excluded the assumption that the user has access to trojaned samples under the threat model. They thereby appear to be ineffective to detect source-label-specific triggers---all these works mainly focus on the commonplace input-agnostic trojan attacks. Detecting source-label-specific triggers, regarded as a challenge, leaves an important future work in the trojan detection research. 


\subsection{Entropy Manipulation}
STRIP examines the entropy of inputs. An attacker might choose to manipulate the entropy of clean and trojaned inputs to eliminate the entropy difference between them. In other words, the attacker can forge a trojaned model exhibiting similar entropy for both clean and trojaned samples. We refer to such an adaptive attack as an entropy manipulation.

An identified specific method to perform entropy manipulation follows the steps below:
\begin{enumerate}
    \item We first poison a small fraction of training samples (specifically, 600) by stamping the trigger c. Then, we (as an attacker) change all the trojaned samples' labels to the attacker's targeted class.
    \item For each poisoned sample, we first randomly select $N$ images (10 is used) from training dataset and superimpose each of $N$ images (clean inputs) with the given poisoned (trojaned) sample. Then, for each superimposed trojaned sample, we randomly assign a label to it and include it into the training dataset.
\end{enumerate}

\begin{figure}[h]
	\centering
	\includegraphics[trim=0 0 0 0,clip,width=0.35\textwidth]{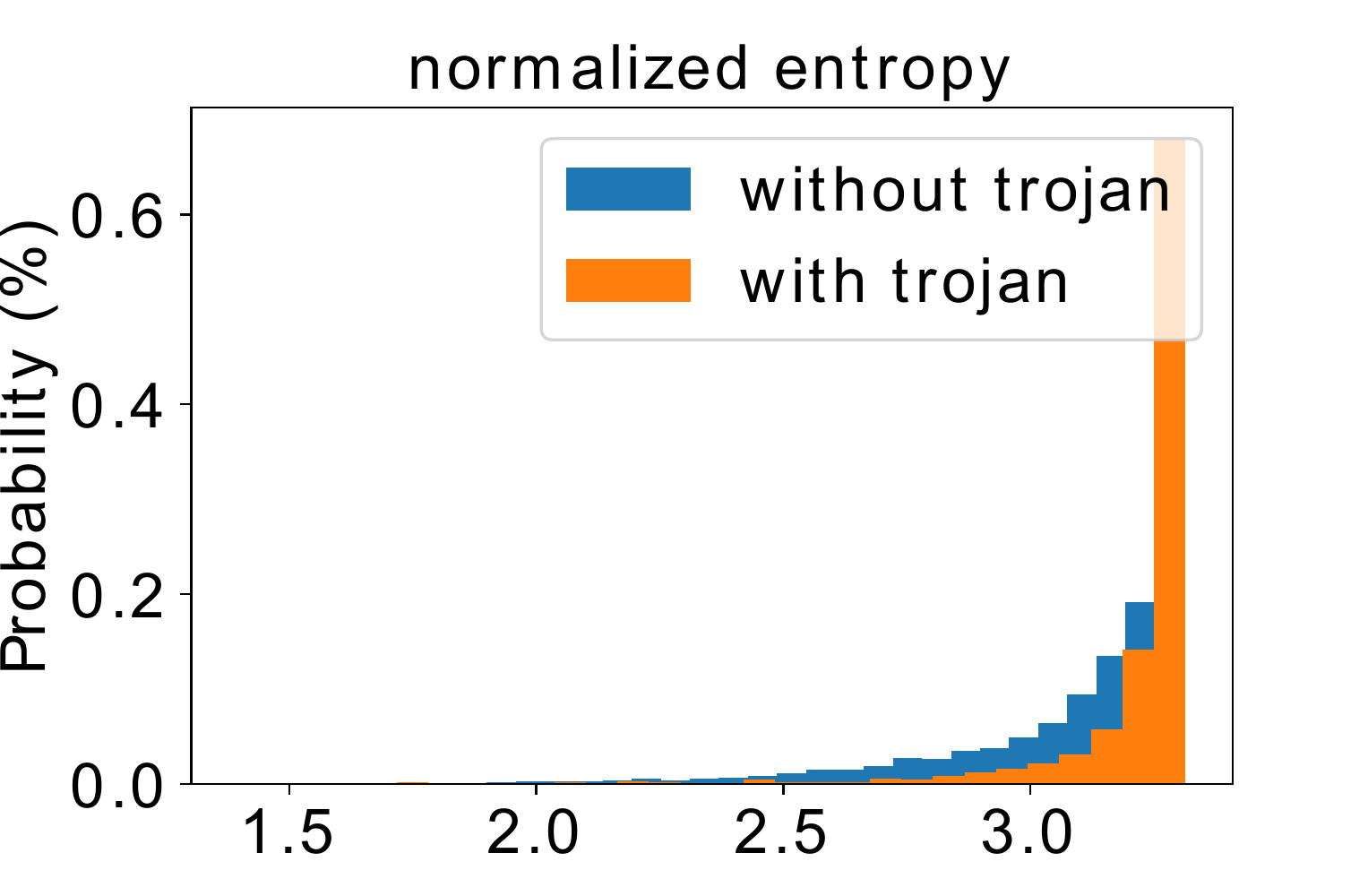}
	\caption{Entropy distribution of clean and trojaned inputs under entropy manipulation adaptive attack. CIFIAR10 and trigger c are used.}
	\label{fig:EnDistTriggerC_AdaptiveAttackDistribution_DNN}
\end{figure}

The intuition of step (2) is to cause predictions of perturbed trojaned inputs to be random and similar to predictions of perturbed clean inputs. After training the trojaned model using the above created poisoned dataset, we found that the classification rate for clean input is 86.61\% while the attack success rate is 99.95\%. The attack success rate drops but is quite small---originally it was 100\% as detailed in Table~\ref{tab:accuracy}. The attacker can still successfully perform the trojan attack. As shown in Fig.~\ref{fig:EnDistTriggerC_AdaptiveAttackDistribution_DNN}, the entropy distribution of clean and trojaned inputs are similar.

However, when the entropy distribution of the clean inputs is examined, it violates the expected {\it normal distribution}~\footnote{We have also tested such an adaptive attack on the GTSRB dataset, and observed the same abnormal entropy distribution behavior of clean inputs.}. In addition, the entropy appears to be much higher. It is always more than 3.0, which is much higher than that is shown in Fig.~\ref{fig:EntropDistAll} (d). Therefore, such an adaptive attack can be detected in practice by examining the entropy of clean inputs (without reliance on trojaned inputs) via the proposed strong perturbation method. Here, the abnormal entropy distribution of the clean inputs indicates a malicious model.


\section{Related Work and Comparison}\label{Sec:Related}
Previous poisoning attacks usually aim to degrade a classifier's accuracy of clean inputs~\cite{huang2011adversarial,papernot2016towards}. In contrast, trojan attacks maintain prediction accuracy of clean inputs as high as a benign model, while misdirecting the input to a targeted class whenever the input contains an attacker chosen trigger.

\subsection{Attacks}
In 2017, Gu {\it et al.}~\cite{gu2017badnets,gu2019badnets} proposed Badnets, where the attacker has access to the training data and can, thus, manipulate the training data to insert an arbitrarily chosen trigger and also change the class labels. Gu {\it et al.}~\cite{gu2017badnets} use a square-like trigger located at the corner of the digit image of the MNIST data to demonstrate the trojan attack. On the MNIST dataset, the authors demonstrate an attack success rate of over 99\% without impacting model performance on benign inputs. In addition, trojan triggers to misdirect traffic sign classifications have also been investigated in~\cite{gu2017badnets}. Chen {\it et al.}~\cite{chen2017targeted} from UC Berkeley concurrently demonstrated such backdoor attacks by poisoning the training dataset. 

Liu~{\it et al.}~\cite{liu2018trojaning} eschew the requirements of accessing the training data. Instead, their attack is performed during the model update phase, not model training phase. They first carry out reverse engineer to synthesize the training data, then improve the trigger generation process by delicately designing triggers to maximize the activation of chosen internal neurons in the neural network. This builds a stronger connection between triggers and internal neurons, thus, requiring less training samples to insert backdoors.

Bagdasaryan~{\it et al.}~\cite{bagdasaryan2018backdoor} show that federated learning is fundamentally vulnerable to trojan attacks. Firstly, participants are enormous, e.g., millions, it is impossible to guarantee that none of them are malicious. Secondly, federated learning is designed to have no access to the participant's local data and training process to ensure the privacy of the sensitive training data; therefore, participants can use trojaned data for training. The authors demonstrate that with controll over no more than 1\% participants, an attacker is able to cause a global model to be trojaned and achieves a 100\% accuracy on the trojaned input even when the attacker is only selected in a single round of training---federated learning requires a number of rounds to update the global model parameters. This federated learning trojan attack is validated through the CIFAR10 dataset that we also use in this paper.

\begin{table*}
	\centering 
	\caption{Comparison with other trojan detection works.}
			\resizebox{0.85\textwidth}{!}{
	\begin{tabular}{l| c | c | c | c | c | c | c} %
		\toprule 
		\toprule 
				
		Work &  \begin{tabular}{@{}c@{}} Black/White  \\ -Box Access$^1$ \end{tabular}  & \begin{tabular}{@{}c@{}} Run-time \end{tabular} & \begin{tabular}{@{}c@{}} Computation  \\ Cost \end{tabular} & \begin{tabular}{@{}c@{}} Time  \\ Overhead \end{tabular} & \begin{tabular}{@{}c@{}} Trigger Size  \\ Dependence \end{tabular} & \begin{tabular}{@{}c@{}} Access to  \\ Trojaned Samples \end{tabular} & \begin{tabular}{@{}c@{}} Detection  \\ Capability \end{tabular} \\ 
		\midrule
		\begin{tabular}{@{}c@{}} Activation Clustering (AC) by  Chen {\it et al.}~\cite{chen2018detecting} \end{tabular}&  White-box & No &  Moderate & Moderate & No & Yes & \begin{tabular}{@{}c@{}} F1 score nearly 100\%\\  \end{tabular}\\ 
		\midrule
		\begin{tabular}{@{}c@{}} Neural Cleanse by Wang {\it et al.}~\cite{wangneural} \end{tabular} &  Black-box & No &  High & High & Yes & No & 100\%$^2$\\
		\hline
		\begin{tabular}{@{}c@{}} SentiNet by Chou {\it et al.}~\cite{chou2018sentinet} \end{tabular} &  Black-box & Yes &  Moderate & Moderate & Yes & No & 5.74\% FAR and 6.04\% FRR\\ 
		\hline		
	    STRIP by us &  Black-box & Yes &  Low & Low & No & No & 0.46\% FAR and 1\% FRR$^3$\\ \hline	
		\bottomrule
	\end{tabular}}
	  \begin{tablenotes}
      \small
      \item $^1$ White-box requires access to inner neurons of the model.
      \item $^2$ According to case studies on 6 infected, and their matching original model, authors~\cite{wangneural} show all infected/trojaned and clean models can be clearly distinguished.
      \item $^3$ The {\it average} FAR and FRR of SentiNet and STRIP are on different datasets as SentiNet does not evaluate on MNIST and CIFAR10. 
    \end{tablenotes}
	\label{tab:comparison} 
\end{table*}
\subsection{Defenses}
Though there are general defenses against poisoning attacks~\cite{baracaldo2017mitigating}, they cannot be directly mounted to guard against trojan attacks. 
Especially, considering that the user has no knowledge of the trojan trigger and no access to trojaned samples, this makes combating trojan attacks more challenging.

Works in~\cite{liu2018fine,liu2017neural} suggest approaches to remove the trojan behavior without first checking whether the model is trojaned or not. Fine-tuning is used to remove potential trojans by pruning carefully chosen parameters of the DNN model~\cite{liu2018fine}. However, this method substantially degrades the model accuracy~\cite{wangneural}. It is also cumbersome to perform removal operations to any DNN model under deployment as most of them tend to be benign. Approaches presented in~\cite{liu2017neural} incur high complexity and computation costs. 

Chen {\it et al.}~\cite{chen2018detecting} propose an activation clustering (AC) method to detect whether the training data has been trojaned or not prior to deployment. The intuition behind this method is that reasons why the trojaned and the benign samples receive same predicted label by the trojaned DNN model are different. By observing neuron activations of benign samples and trojaned samples that produce same label in hidden layers, one can potentially distinguish trojaned samples from clean samples via the activation difference. This method assumes that the user has access to the trojaned training samples in hand.

Chou {\it et al.}~\cite{chou2018sentinet} exploit both the model interpretability and object detection techniques, referred to as SentiNet, to firstly discover contiguous regions of an input image important for determining the classification result. This region is assumed having a high chance of possessing a trojan trigger when it strongly affects the classification. Once this region is determined, it is carved out and patched on to other held-out images that are with ground-truth labels. If both the misclassification rate---probability of the predicted label is not the ground-truth label of the held-out image---and confidence of these patched images are high enough, this carved patch is regarded as an adversarial patch that contains a trojan trigger. Therefore, the incoming input is a trojaned input. 

In Oakland 2019, Wang {\it et al.}~\cite{wangneural} propose the Neural Cleanse method to detect whether a DNN model has been trojaned or not prior to deployment, where its accuracy is further improved in~\cite{guo2019tabor}. Neural Cleanse is based on the intuition that, given a backdoored model, it requires much smaller modifications to all input samples to misclassify them into the attacker targeted (infected) label than any other uninfected labels. Therefore, their method iterates through all labels of the model and determines if any label requires a substantially smaller amount of modification to achieve misclassifications.
One advantage of this method is that the trigger can be discovered and identified during the trojaned model detection process. However, this method has two limitations. 
Firstly, it could incur high computation costs proportionally to the number of labels. 
Secondly, similar to SentiNet~\cite{chou2018sentinet}, the method is reported to be less effective with increasing trigger size.

\vspace{-0.3cm}
\subsection{Comparison}\label{sec:limitation}
We compare STRIP with other three recent trojan detection works, as summarized in Table~\ref{tab:comparison}. Notably, AC and Neural Cleanse are performed offline prior to the model deployment to {\it directly detect whether the model has been trojaned or not}. In contrast, SentiNet and STRIP are undertake run-time checking of incoming inputs to {\it detect whether the input is trojaned or not when the model is actively deployed}. STRIP is efficient in terms of computational costs and time overhead. While AC and STRIP are insensitive to trojan trigger size, AC assumes access to a trojaned sample set.

We regard SentiNet to be mostly related to our approach since both SentiNet and STRIP focus on detecting whether the incoming input has been trojaned or not during run-time. However, there are differences: i) We do not care about the ground-truth labels of neither the incoming input nor the drawn images from the held-out samples, while~\cite{chou2018sentinet} relies on the ground-truth labels of the held-out images; ii) We introduce entropy to evaluate the randomness of the outputs---this is more convenient, straightforward and easy-to-implement in comparison with the evaluation methodology presented in~\cite{chou2018sentinet}; iii) STRIP evaluations demonstrate its capability of detecting a large trigger. One limitation of SentiNet is that the region embedding the trojan trigger needs be small enough. If the trigger region is large, such as the trigger shown in Fig.~\ref{fig:Trigger} (a) and (c), and Fig.~\ref{fig:largeTriggerExam}, then SentiNet tends to be less effective. This is caused by its carve-out method. 
Supposing that the carved region is large and contains the trigger, then patching it on held-out samples will also show a small misclassification rate to be falsely accepted as a benign input via SentiNet. 

Notably, in contrast to the use of a global detection boundary in Neural Cleanse~\cite{wangneural}, the detection boundary of STRIP is unique to each deployed model and is {\it extracted from the already deployed model itself}; this boundary is not a global setting. This avoids the potential for the global setting to fail since the optimized detection boundary for each model can vary. 
Probably, one not obvious fact is that users need to train trojan/clean models by themselves to find out this global setting as the detection boundary of the Neural Cleanse needs to be decided based on reference models---STRIP does not need reference model but solely the already deployed (begin/backdoored) model. This may partially violate the motivation for outsourcing the model training of ML models---the main source of attackers to introduce backdoor attacks: if the users own training skills and the computational power, it may be reasonable to train the model, from scratch, by themselves.

\vspace{-0.3cm}
\subsection{Watermarking}
There are works considering a backdoor as a watermark~\cite{chen2018blackmarks} to protect the intellectual property (IP) of a trained DNN model~\cite{adi2018turning,guo2018watermarking,zhang2018protecting}. The argument is that the inserted backdoor can be used to claim the ownership of the model provider since only the provider is supposed to have the knowledge of such a backdoor, while the backdoored DNN model has no (or imperceptible) degraded functional performance on normal inputs. However, as the above countermeasures---detection, recovery, and removal---against backdoor insertion are continuously evolved, the robustness of using backdoors as watermarks is potentially challenged in practical usage. We leave the robustness of backdoor entangled watermarking under the backdoor detection and removal threat as part of future work since it is out of the scope of this work.

\vspace{-0.3cm}
\section{Conclusion and Future Work}
The presented STRIP constructively turns the strength of insidious input-agnostic trigger based trojan attack into a weakness that allows one to detect trojaned inputs (and very likely backdoored model) at run-time.
Experiments on MNIST, CIFAR10 and GTSRB datasets with various triggers and evaluations validate the high detection capability of STRIP. Overall, the FAR is lower than 1\%, given a preset FRR of 1\%. The 0\% FRR and 0\% FAR are empirically achieved on popular CIFAR10 and GTSRB datasets. While easy-to-implement, time-efficient and complementing with existing trojan mitigation techniques, the run-time STRIP works in a black-box manner and is shown to be capable of overcoming the trigger size limitation of other state-of-the-art detection methods. Furthermore, STRIP has also demonstrated its robustness against several advanced variants of input-agnostic trojan attacks and the entropy manipulation adaptive attack.

Nevertheless, similar to Neural Cleanse~\cite{wangneural} and SentiNet~\cite{chou2018sentinet}, STRIP is not effective to detect source-label-specific triggers; this needs to be addressed in future work. In addition, we will test STRIP's generalization to other domains such as text and voice . 

%




\appendices
\section{Trigger Transparency Results}\label{sec:appendTrans}
\begin{figure}[h!]
	\centering
	\includegraphics[trim=0 0 0 0,clip,width=0.50\textwidth]{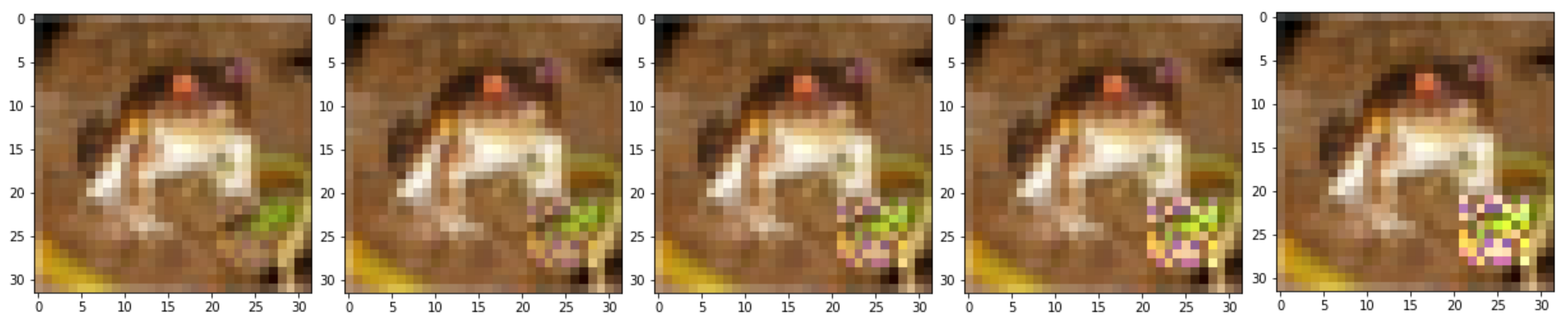}
	\caption{From left to right, trigger transparency are 90\%, 80\%, 70\%, 60\% and 50\%.}
	\label{fig:transp}
\end{figure}

Fig.~\ref{fig:transp} shows different transparency settings. Table~\ref{tab:transp} details classification rate of clean inputs, attack success rate of trojaned inputs, and detection rate under different transparency settings.

\begin{table}[h!]
	\centering 
	\caption{Classification rate of clean images, attack success rate and detection capability under different trigger transparency settings. Dataset is CIFAR10 and the trigger is trigger b in Fig.~\ref{fig:Trigger} (b). The FRR is preset to be 0.5\%.}			\resizebox{0.5\textwidth}{!}{
	\begin{tabular}{c| c | c | c | c | c | c} %
		\toprule 
		\toprule 
				
		Transp. &  \begin{tabular}{@{}c@{}} Classification rate  \\ of clean image \end{tabular}  & \begin{tabular}{@{}c@{}} Attack  \\ success rate \end{tabular} & \begin{tabular}{@{}c@{}} Min. entropy \\ of clean images \end{tabular} & \begin{tabular}{@{}c@{}} Max. entropy \\ of trojaned images \end{tabular} & \begin{tabular}{@{}c@{}} Detection \\ boundary \end{tabular} & \begin{tabular}{@{}c@{}} FAR \end{tabular} \\ 
		\midrule
		90\% &  87.11\% & 99.93\% &  0.0647 & 0.6218 & 0.2247 & 0.10\% \\ 
		\midrule
		80\% &  85.81\% & 100\% &  0.0040 & 0.0172 & 0.1526 & 0\% \\ \hline
		70\% &  88.59\% & 100\% &  0.0323 & 0.0167 & 0.1546 & 0\% \\ \hline
		60\% &  86.68\% & 100\% &  0.0314 & $3.04\times 10^{-17}$ & 0.1459 & 0\% \\ \hline	
		50\% &  86.80\% & 100\% &  0.0235 & $4.31\times 10^{-6}$ & 0.1001 & 0\% \\ 		
		\bottomrule
	\end{tabular}
			}
	\label{tab:transp} 
\end{table}

\section{Detection Capability Relationship with Depth of Neural Network}\label{App:depth}
\begin{figure}[h!]
	\centering
	\includegraphics[trim=0 0 0 0,clip,width=0.4\textwidth]{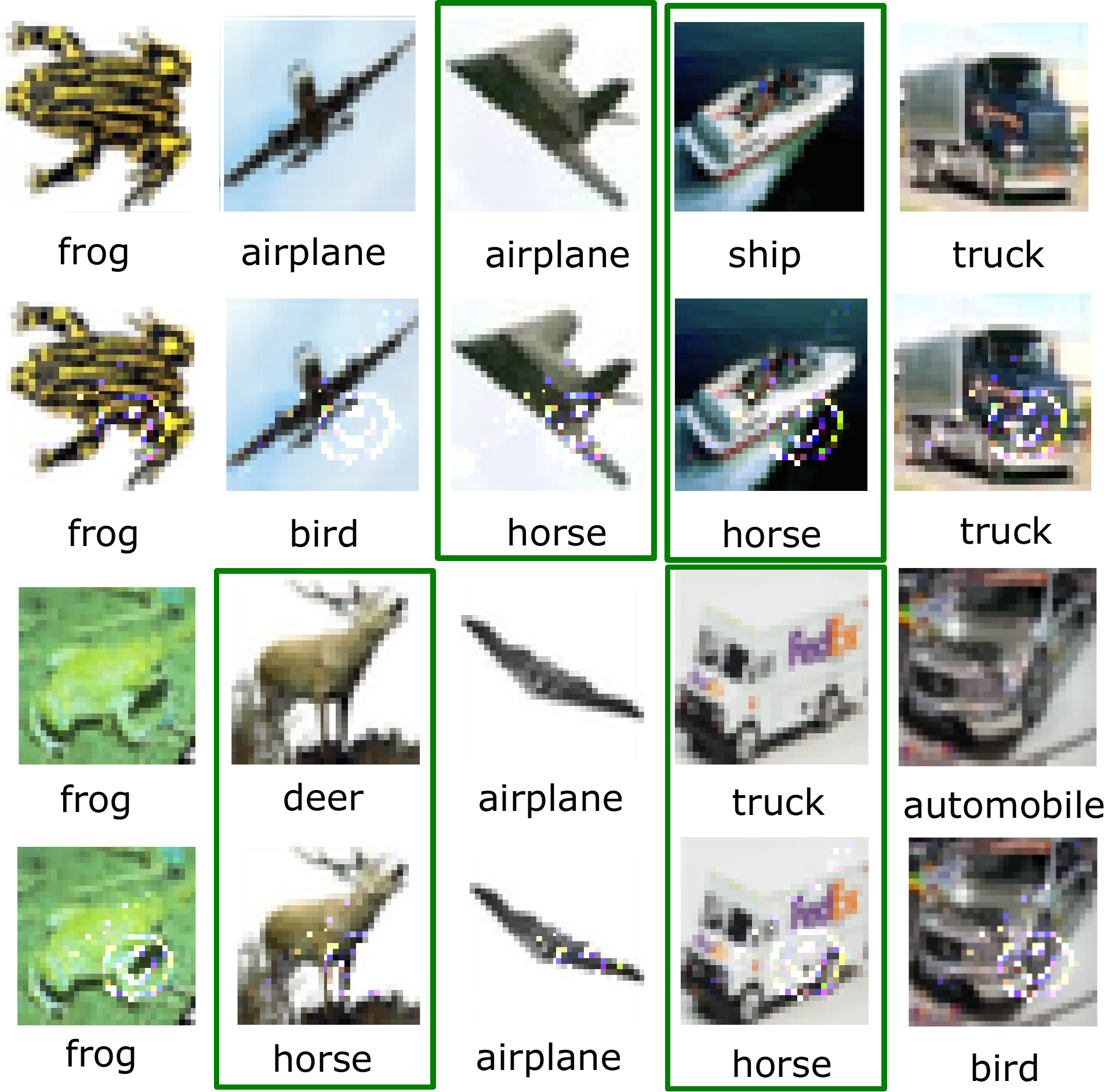}
	\caption{When the trojaned images are falsely accepted by STRIP as benign images, most of them lost their trojaning effect. Because they cannot hijack the trojaned DNN model to classify them to the targeted class---`horse'. Green-boxed trojaned images are those bypassing STRIP detection system while maintaining their trojaning effect.}
	\label{fig:falseAccepted}
\end{figure}
Besides the DNN architecture---referred to as 8-layer architecture---achieving around 88\% accuracy performance of clean inputs, we tested a shallow neural network architecture only with 2 conventional layer and 1 dense layer---referred to as 2-layer architecture. For this 2-layer architecture, the benign model on CIFAR10 dataset has a lower accuracy performance, which is 70\%. The corresponding trojaned model with trigger c has a similar accuracy with around 70\% for clean inputs while around 99\% attack success rate for trojaned inputs. In this context, the model is successfully inserted as it does not degrade the performance of clean inputs. 

We find that as the neural network goes deeper---usually leads to a more accurate prediction, the detection capability also improves. Specifically, for the shallow 2-layer architecture based trojaned model, 2\% FRR gives 0.45\% FAR, 1\% FRR gives 0.6\% FAR, and 0.5\% FRR gives 0.9\% FAR. While for the 8-layer architecture based trojaned model, FRR is always 0\%, regardless of FRR, as there is always an entropy gap---no overlap---between the benign and trojaned inputs. 

Moreover, we run a 8-layer architecture on the MNIST dataset with the square trigger. For the trojaned model, its accuracy on clean inputs is 99.02\% while achieves a 99.99\% accuracy on trojaned inputs. STRIP demonstrates an improved detection capability as well. Specifically, 1\% FRR gives 0\% FAR, 0.5\% FRR gives 0.03\% FAR, which has been greatly improved in comparison with the detection capability of a 2-layer trojaned model, see Table.~\ref{tab:FRRFAR}.

To this end, we can empirically conclude that the deeper the model, the higher detection capability of STRIP detection. On one hand, this potentially lies on the fact that the model with more parameters memorizes the trigger feature stronger, which always presents a low entropy for the trojaned input. On the other hand, the model also more accurately memorizes the features for each class of clean input. The trained model is more sensitive to strong perturbation on clean input, and therefore, unlikely to present a low entropy for clean input---may contribute to FRR.

We are curious on those images that are trojaned but falsely accepted as clean images. Therefore, based on the 2-layer trojaned model (8-layer model has 0\% FAR) produced on the CIFAR10 dataset and trigger c, we further examined those images. We found that most of them lost their trojan effect, as shown in Fig.~\ref{fig:falseAccepted}. For instance, out of 10 falsely accepted trojaned images, four images maintaining their trojaning effect of hijacking the DNN model to classfy them to be the targeted label of `horse'. The rest six trojaned images are unable to achieve their trojaning effect because the trojan trigger is not strong enough to misdirect the predicted label to be `horse'. In other words, these six trojaned images will not cause security concerns {\it designed by the attacker} when they are indeed misclassified into benign image by STRIP. In addition, we observe that there are three trojaned images classified into their correct ground-truth labels by the attacker's trojaned model. The reason may lie on that the trigger feature is weakened in certain specific inputs. For example, without careful attention, one may not perceive the stamped trigger in the `frog' (1st) and `airplane' (7th) images, which is more likely the same to the trojaned DNN model.

\end{document}